\documentclass{article}

\usepackage[preprint,nonatbib]{neurips_2022}

\usepackage[citestyle=numeric-comp,sorting=none,maxbibnames=99,url=false,doi=false,giveninits=true]{biblatex}
\DeclareFieldInputHandler{title}{}

\usepackage[utf8]{inputenc} 
\usepackage[T1]{fontenc}    
\usepackage{hyperref}       
\usepackage{url}            
\usepackage{booktabs}       
\usepackage{amsfonts}       
\usepackage{nicefrac}       
\usepackage{microtype}      
\usepackage[table]{xcolor}         
\usepackage[pdftex]{graphicx} 
\usepackage{threeparttable} 
\usepackage{booktabs}       
\usepackage{array}
\usepackage{caption}
\usepackage{subcaption}
\usepackage{pifont}

\usepackage[longend]{algorithm2e}
\usepackage{textgreek}
\usepackage{amssymb}
\usepackage[version=4]{mhchem}
\usepackage{tcolorbox}

\usepackage{amsmath}
\usepackage{bm}

\title{Photonics for Neuromorphic Computing: Fundamentals, Devices, and Opportunities}

\author{
  \small Renjie Li \textsuperscript{$1$} 
  \And
  \small Yuanhao Gong \textsuperscript{$1$} 
  \And
  \small Hai Huang \textsuperscript{$1$} 
  \And
  \small Yuze Zhou \textsuperscript{$1$} 
  \And
  \small Sixuan Mao \textsuperscript{1} 
  \And
  \small Zhijian Wei \textsuperscript{$*,2$} \: \: \:
  \small Zhaoyu Zhang \textsuperscript{$*,1$} \\
  \small \texttt{Email: zhangzy@cuhk.edu.cn} \\
\tiny \textsuperscript{1} School of Science and Engineering, Guangdong Key Laboratory of Optoelectronic Materials and Chips, \\ \tiny Shenzhen Key Lab of Semiconductor Lasers, The Chinese University of Hong Kong, Shenzhen, Guangdong, China \\
\tiny \textsuperscript{2}  SONT Technologies Co., LTD, Shenzhen, China \\
\small \textsuperscript{*} indicates corresponding authors
}
\begin{document}

\maketitle
\begin{abstract}
In the dynamic landscape of Artificial Intelligence (AI), two notable phenomena are becoming predominant: the exponential growth of large AI model sizes and the explosion of massive amount of data. Meanwhile, scientific research such as quantum computing and protein synthesis increasingly demand higher computing capacities. Neuromorphic computing, inspired by the mechanism and functionality of human brains, uses physical artificial neurons to do computations and is drawing widespread attention. Conventional electronic computing has experienced certain difficulties, particularly concerning the latency, crosstalk, and energy consumption of digital processors. As the Moore's law approaches its terminus, there is a urgent need for alternative computing architectures that can satisfy this growing computing demand and break through the von Neumann model. Recently, the expansion of optoelectronic devices on photonic integration platforms has led to significant growth in photonic computing, where photonic integrated circuits (PICs) have enabled ultrafast artificial neural networks (ANN) with sub-nanosecond latencies, low heat dissipation, and high parallelism. Such non-von Neumann photonic computing systems hold the promise to cater to the escalating requirements of AI and scientific computing.  In this review, we study recent advancements in integrated photonic neuromorphic systems, and from the perspective of materials and device engineering, we lay out the scientific and technological breakthroughs necessary to advance the state-of-the-art. In particular, we examine various technologies and devices employed in neuromorphic photonic AI accelerators, spanning from traditional optics to PICs. We evaluate the performances of different designs by energy efficiency in operations per joule (OP/J) and compute density in operations per squared millimeter per second (OP/$mm^2$/s). Putting special emphasis on photonic components such as VCSEL lasers, optical interconnects, and frequency microcombs, we highlight the most recent breakthroughs in photonic engineering and materials science used to create advanced neuromorphic computing chips. Lastly, we recognize that existing technologies encounter obstacles in achieving photonic AI accelerators with peta-level computing speed and energy efficiency, and we also explore potential approaches in new devices, fabrication, materials, and integration to drive innovation. As the current challenges and barriers in cost, scalability, footprint, and computing capacity are resolved one-by-one, photonic neuromorphic systems are bound to co-exist with, if not replace, conventional electronic computers and transform the landscape of AI and scientific computing in the foreseeable future.

\end{abstract}

\newpage
\tableofcontents
\newpage

\section{Introduction}

In the dynamic landscape of Artificial Intelligence (AI) \cite{russel2010}, two notable trends have shaped its development: the exponential growth in the number of AI model parameters and the staggering amount of data generated. These factors have propelled AI to new heights and unlocked unprecedented possibilities. The first trend revolves around the expansion of AI model parameters (Figure 1). In recent years, there has been a remarkable surge in the size and complexity of deep neural networks (DNN) (Figure 2b). Deep learning models have shown remarkable capabilities in tasks such as image classification, natural language processing, and speech recognition. The increase in the number of parameters within these large deep learning models has played a significant role in their success because by augmenting its capacity to learn intricate patterns and representations, larger models can often achieve superior performance. Large language models (LLM) \cite{wei2022emergent}, such as chatGPT for example, have achieved unprecedented scale and popularity since April 2023. It's reported that GPT-3 and GPT-4 consists of approximately 175 billion and 1.8 trillion parameters, respectively. These parameters serve as the variables that the model learns and adapts from training data, enabling it to generate coherent and contextually relevant text responses from prompts. This remarkable expansion has been made possible by advancements in computational hardware, such as Graphics Processing Units (GPUs) and specialized hardware accelerators like Google's Tensor Processing Unit (TPU), which enable the efficient training and inference of such complex models. Simultaneously, the second trend highlights the explosion of data generation. In the digital era, it is estimated that around 2.5 quintillion bytes (2.5 exabytes, exa=$10^{18}$) of data are created daily, and over the next three years up to 2025, global data creation is projected to accumulate to more than 180 zettabytes (zetta=$10^{21}$). From social media interactions and online transactions to sensor measurements and scientific research, this data serves as a rich resource for training and fine-tuning AI models. The availability of vast and diverse datasets has in turn revolutionized AI. Data-driven approaches, often referred to as "big data", have enabled AI models to learn from enormous amounts of information, leading to improved accuracy and robustness. Additionally, advancements in data storage, processing, and cloud technologies have made it easier to handle and analyze massive datasets efficiently. The increased capacity of large models to capture and represent complex relationships, coupled with the abundance of training data, has facilitated breakthroughs in diverse domains, including healthcare, finance, transportation, robotics, metaverse, and natural language understanding.

\begin{figure}[hbt!]
\centering
\includegraphics[width=1.0\textwidth]{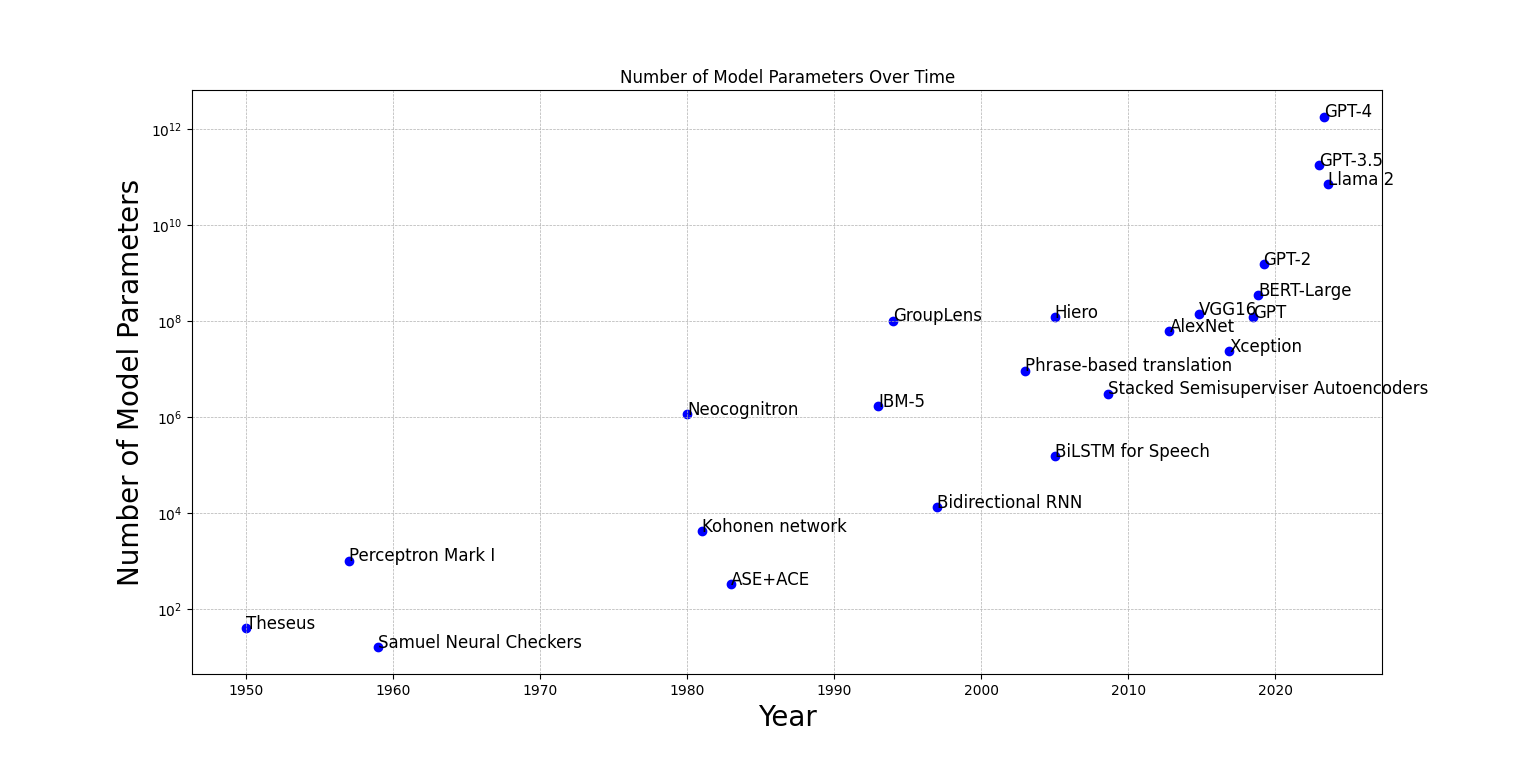}
\caption{Trend of increase in representative AI model's parameters over time. Parameters (i.e. weights) are variables in an Al system whose values are adjusted during training to establish how input data is correlated with the output labels. The most recent chatGPT has a record-breaking 1.8 trillion parameters and is still growing. No. of parameters are estimated based on published statistics in respective papers and naturally come with some uncertainty. }\label{fig:1}
\index{figures}
\end{figure}

However, this growth is not without challenges. The large number of model parameters and the amount of data required for training pose significant computational and resource-intensive requirements. Developing new chip hardware and computing paradigms to scale AI models efficiently, along with addressing concerns related to latency, parallelism, and energy consumption, are areas that researchers and engineers are actively exploring.

Neuromorphic computing \cite{tang2019bridging,upadhyay2019emerging,torres2023thermal,schuman2022opportunities,shastri2021photonics}, an emerging field of research, focuses on developing computational systems inspired by the structural and functional characteristics of the human brain. This discipline strives to overcome the limitations of conventional computing by mimicking the parallelism, fault tolerance, and energy efficiency observed in biological neural networks (Figure 2a). At its core, neuromorphic computing aims to create hardware and software architectures that replicate the behavior of neurons and synapses. These architectures enable the processing of information using spiking neural networks and specialized neuromorphic chips, which offer real-time handling of complex data and the potential for accelerated machine learning algorithms. The primary advantage of neuromorphic computing lies in its ability to process information in a massively parallel manner, leading to enhanced computational and energy efficiency. By leveraging the inherent capabilities of neurons and synapses, neuromorphic systems demonstrate potential applications in areas such as pattern recognition, image classification, autonomous driving, and LLM. In hardware design, researchers are exploring novel components like micro-ring resonators and VCSEL lasers to emulate synaptic connections or to perform matrix operations. These photonic components enable the creation of highly efficient AI hardware that can adapt and learn from input data, similar to the plasticity observed in biological neurons. Neuromorphic techniques enable the development of self-learning systems capable of processing sensory input and making intelligent decisions in dynamic environments to facilitate the growth of AI.

\begin{figure}[hbt!]
\centering
\includegraphics[width=.75\textwidth]{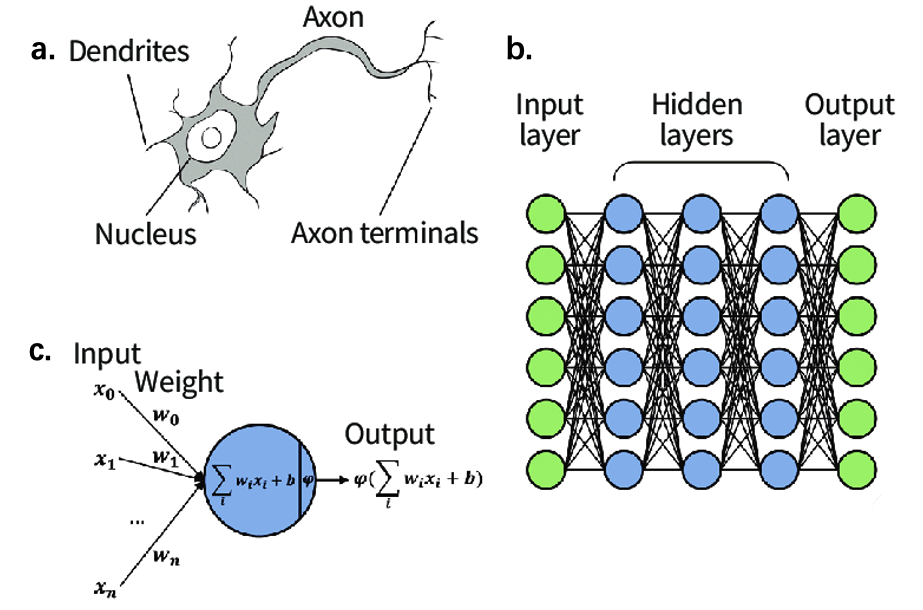}
\caption{a. Biological neuron in animals; b. Multi-layer perception neural networks (MLP) or fully-connected (FC) layers; c. Forward propagation of artificial neurons in MLP, including the input, weighs, summation, activation function, and the output. Data obtained from online image libraries.}\label{fig:1}
\index{figures}
\end{figure}

In optical computing, photonic devices that utilize optical near-fields and effective interactions are very important. Key components within neural networks (Figure 2b-c), such as activation functions \cite{williamson2019reprogrammable, Miscuglio_Mehrabian_Hu_Azzam_George_Kildishev_Pelton_Sorger_2018, Mourgias-Alexandris_Tsakyridis_Passalis_Tefas_Anastasios, Zuo_Li_Zhao_Jiang_Chen_Chen_Jo_Liu_Du_2019} and backpropagation \cite{hughes2018training}, currently represent the focal point of research in the realization of optical neural networks.
Y. Zuo et. al. demonstrated an all-optical neural network (AONN) which can work with nonlinear activation \cite{Zuo_Li_Zhao_Jiang_Chen_Chen_Jo_Liu_Du_2019}. They used spatial light modulators and Fourier lenses to realize programmed linear operations and laser-cooled atoms with electromagnetically induced transparency to realize nonlinear activation function. In this device, linear operations are done via Fourier transform and all diffracted beams in the same direction are summed onto a spot, which can be expressed as $z_i=b_i+\Sigma_jW_{ij}x_j$ (Figure 2c). In this formula, \textit{W} is weights, \textit{x} is input, and \textit{b} is bias.  When nonlinear operations are also done, we can get the final output $y_i=\varphi\left(z_i\right)$, where $\varphi$ is a nonlinear activation function (Figure 2c). It is obvious that this neural network which uses Fourier lenses is easy to construct, low in power consumption, and boasts rapid computational speed. However, as a neural network system, the Fourier optical lens system might be excessively bulky.

Another all-optical neural network, termed "D2NN'', is considered as a potential contender for miniaturized all-optical diffractive deep neural networks  \cite{Lin_Rivenson_Yardimci_Veli_Luo_Jarrahi_Ozcan_2018}. In this system, each point on a specified layer is a neuron, and has a designed transmission or reflection coefficients. When it works, the input of each layer is defined by previous layer based on free-space diffraction. In this work, the device is on the macro-scale as it spans a few centimeters in length and width, and a few millimeters in thickness. However, this is because it operates in the terahertz range, and if altered to function in the near-infrared range suitable for optical communications, its size could potentially be reduced by a few hundred times. The fabrication of such tiny devices currently presents a significant challenge, yet D2NN remains a viable contender for integrated AONN systems.

With its rapid development, relative maturity, and high degree of integration, silicon photonics emerges as a compelling candidate for the integration of AONN. In silicon photonic integrated circuit, the Mach-Zehnder Interferometer (MZI) can regulate the intensity of optical signals through light interference. By amalgamating multiple MZIs, we can form an integrated optical neural network (ONN) on-chip. In 2017, Y. Shen et. al. proposed an AONN based on cascaded array of 56 programmable MZI in a silicon photonic integrated circuit \cite{Shen_Harris_Skirlo_Prabhu_Baehr-Jones_Hochberg_Sun_Zhao_Larochelle_Englund_et}. By programming the internal and external phase shifters of each MZI, this system enables arbitrary SU(4) rotations and finally expresses unitary matrices. The splitting ratio and the differential output phase are controlled when the operations between unitary matrices are done. In 2019, I. Williamson et. al. added nonlinear activation function operating on MZI phase shift and realized nonlinear activation function on on-chip AONN \cite{williamson2019reprogrammable}. They improved test accuracy on the MNIST task from 85\% to 94\%.

With earlier development of nonlinear activation functions, AONNs, and D2NNs that serve as the building blocks of photonic neuromorphic computing, researchers later began to explore photonic devices such as frequency microcombs \cite{feldmann2021parallel}, micro-ring resonators \cite{xu202111}, lasers \cite{chen2023deep}, metasurfaces \cite{liu2022programmable}, optical attenuators \cite{ashtiani2022chip}, and photodiodes \cite{ashtiani2022chip} etc. to realize more advanced neuromorphic functionalities. We can roughly categorize existing efforts into three branches: 1. simulate the principles of forward propagation in an artificial neuron; 2. achieve image classification or pattern recognition; 3. realize convolutional operations by performing matrix-vector multiplications (MVM).
A full timeline of the evolution of photonic neuromorphic systems is shown in Figure 3.

\begin{figure}[hbt!]
\centering
\includegraphics[width=.95\textwidth]{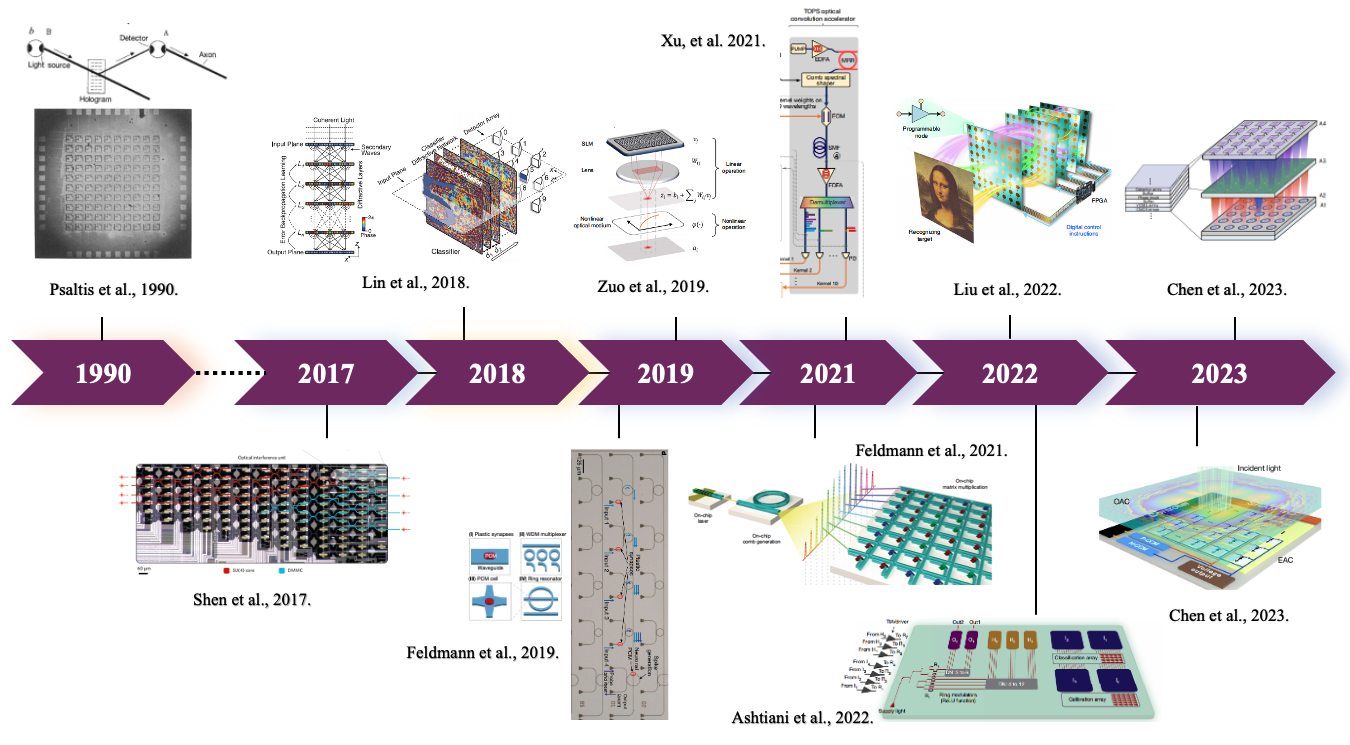}
\caption{Timeline of the evolution of photonic neuromorphic systems, especially for AI applications. Existing works can be categorized into three branches: 1. simulate the principles of forward propagation in an artificial neuron; 2. achieve image classification or pattern recognition; 3. realize convolutional operations by performing MVM. Sources are from milestone works spanning from the 90s to 2023: \cite{psaltis1990holography,Shen_Harris_Skirlo_Prabhu_Baehr-Jones_Hochberg_Sun_Zhao_Larochelle_Englund_et,Lin_Rivenson_Yardimci_Veli_Luo_Jarrahi_Ozcan_2018,feldmann2019all,Zuo_Li_Zhao_Jiang_Chen_Chen_Jo_Liu_Du_2019,feldmann2021parallel,xu202111,ashtiani2022chip} \cite{liu2022programmable,chen2023all,chen2023deep}. For detailed specs of each work, refer to Table 1.}\label{fig:1}
\index{figures}
\end{figure}

About a dozen reviews have been written on the topic of photonic neuromorphic computing systems. Some place heavier emphasis on a specific aspect of the system such as spike generation, synapses or nonlinearity, and some focus on specific structures such as MZI, frequency combs, or diffractive layers, and some others tend to discuss the relationship between photonic neural networks and machine learning via the bridge of multiply-accumulate operations (MAC) and wavelength division multiplexing (WDM).  This review, however, aims to concisely and comprehensively unify each of the aforementioned aspects of photonic neuromorphic design and dissect them from the perspective of photonic engineering and materials science; as a result, details/fundamentals of the theories of neural networks or computing systems will not be fully addressed here. According to Figure 4, the review is organized as follows: Sec. 1 Introduction. Sec. 2 Outlines the history and challenges of neuromorphic computing based on conventional electronics paradigm. Sec. 3 Introduces fundamentals of photonics, including theories and principles of light-matter interaction.  Sec. 4 Covers existing photonic components and fab platforms for neuromorphic computing. Sec. 5 Details recent advances of photonic neuromorphic computing, providing a summary of the evolution of photonic neuromorphic paradigms (also graphed in Figure 3). Sec. 6 Addresses emerging technologies in photonics for neuromorphic computing, including topological insulators and PCSEL lasers. Sec. 7 Discusses the challenges and future directions in photonic neuromorphic computing. Finally, Sec. 8 summarizes the review by providing a conclusion and closing marks.

\begin{figure}[hbt!]
\centering
\includegraphics[width=.75\textwidth]{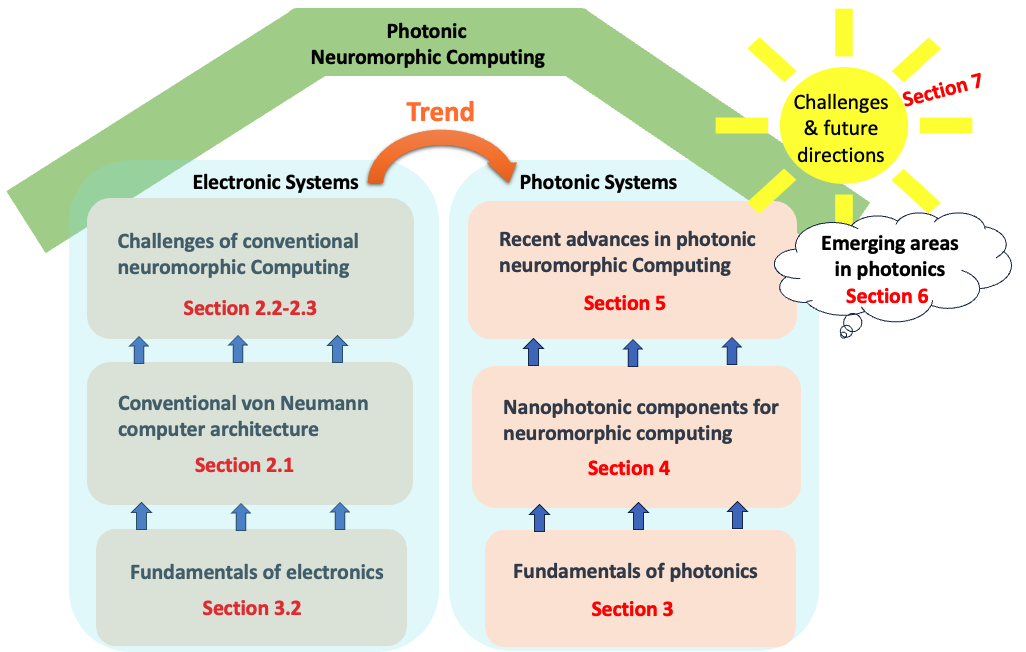}
\caption{Overall hierarchical organization of this paper and logical links between all  sections.}\label{fig:1}
\index{figures}
\end{figure}


\section{Conventional Neuromorphic Computing: History and Challenges}
\subsection{Brief Overview of Conventional Computing and Its Limitations}

With the development of electronic device and improvement of the computing performance, many changes taken place in human society, and many scientific fields have made numerous breakthroughs. In order to meet the needs of scientific development, the size of electronic device is required to continue to scale to support the continuous growth of computing performance and maintain this increasing trend \cite{zidan2018future}. In particular, nowadays, because of the fast-paced advancements of AI, LLMs, Internet of Things (IoT), metaverse, and cloud computing, the amount of data has virtually exploded and the demand for high-performance memory and computing efficiency has become higher and higher \cite{tang2019bridging,huang2020memory}. However, there are currently some difficulties that need to be addressed urgently. On one hand, as the Moore’s law comes to its end, the required performance gains can no longer be achieved through conventional scaling device \cite{agwa5digital,zidan2018future}. On the other hand, at present, most computers that are general-purpose devices are designed based on the traditional Von Neumann architecture. Limited by the bottlenecks of hardware fabrication technology and the inherent structural problems such as “memory wall”, conventional von Neumann architecture can not sustain processing large amounts of data in the AI era \cite{agwa5digital,upadhyay2019emerging}. So it is critical to propose alternative architectures that scale beyond von Neumann to break through the computing bottlenecks. Figure 5 plots figure of merit (FoM) of existing digital electronic hardware and emerging photonic neuromorphic systems as AI accelerators. We evaluate the performances of different architectures by energy efficiency in terms of operations per joule (OP/J) and compute density in terms of operations per squared millimeter per second (OP/$mm^2$/s). By comparison, lab-tested photonic neuromorphic systems already outperform commercial electronic hardware in both energy efficiency and compute density as of February 2024.

\begin{figure}[hbt!]
\centering
\includegraphics[width=.8\textwidth]{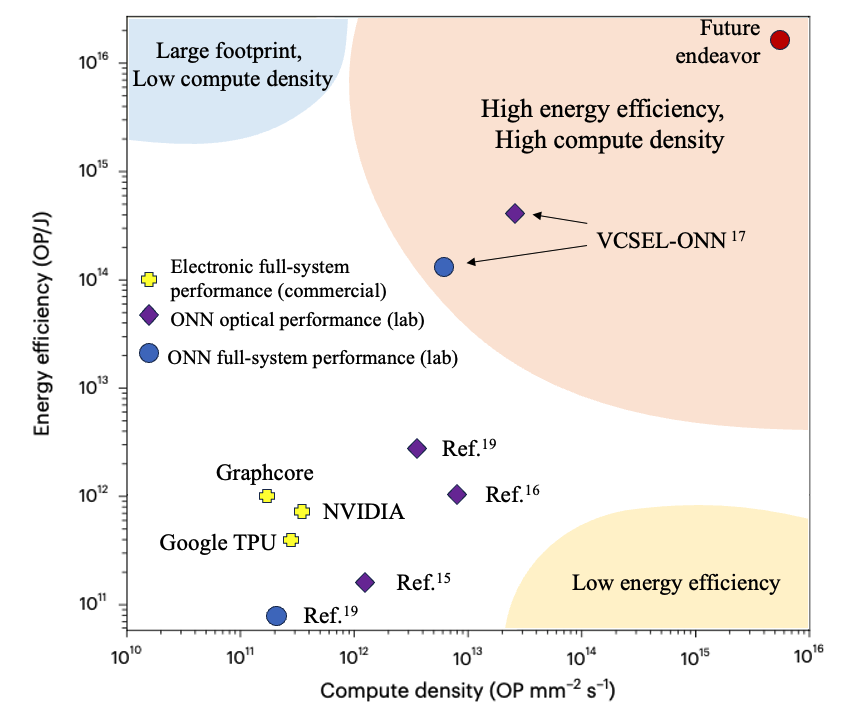}
\caption{Comparison of state-of-the-art AI accelerators between conventional digital electronics and the emerging ONNs empowered by photonic neuromorphics. Comparison is embodied by the FoM such as energy efficiency ($TOP/J$) and compute density ($TOP/mm^2/s$). Corresponding FoM values of each reference can be found in Table 1. On both the x and y axis, larger values are better. As indicated by the legend, it should be noted that while the electronic systems shown here are commercial products, the photonic (ONN) systems are laboratory demonstrations. Future endeavor calls for continued efforts to improve the energy efficiency and compute density of photonic systems. Adapted with permission.\cite{chen2023deep} Copyright 2023, Springer Nature.}\label{fig:1}
\index{figures}
\end{figure}

\subsubsection{Brief Overview of Conventional Computing}

Progress in the computing domain has significantly relied on semiconductor devices' miniaturization. In 1965, Intel cofounder Gordon Moore paid attention to the stable growth rate of miniaturization and published a prediction known as “Moore’s Law”, which states that the number of transistors in each new-generation of computer chips would double every two years \cite{leiserson2020there}. With the miniaturization of transistors, the density of transistors in chips increases continuously, and the computing capability is also becoming stronger and stronger. Moore’s Law has driven progress in many aspects of computing, such as better CPU performance, improved energy efficiency, larger storage capacity and better cost savings \cite{kim2017heterogeneous}. Moreover, a computer architecture named the “Von Neumann Architecture” revolutionized computing as Moore’s Law took effect, and enabled hardware engineers to build a variety of computational systems \cite{upadhyay2019emerging}. 

Von Neumann Architecture, also known as the Princeton Architecture, is the fundamental organizational structure of a digital computer based on the principles proposed by mathematician John von Neumann. It treats instructions (the computer program) as a special type of data and stores instructions and data in different addresses in the same memory. The main characteristics of the Von Neumann architecture is that it adopts a binary system and computations execute in a procedural order. The invention of this type of architecture laid the foundation for modern computer architectural concepts \cite{xue2021overview}. 

Existing digital computers are built upon the Von Neumann architecture and based on the silicon microelectronics platform. Generally, a Von Neumann computer is mainly composed of a memory bank for storing data and instructions and a central processing unit (CPU) for performing nonlinear operations and connecting transmissions between the two \cite{nahmias2019photonic,backus1978can,schuman2022opportunities}. Among them, as the core of the architecture, the CPU is composed of a control unit and an arithmetic logic unit (ALU). Usually, the CPU executes a series of instructions to process data stored in memory units through interacting with the memory system \cite{arikpo2007neumann}. Main memory is the key to memory system and set up by adopting the Dynamic Random Access Memory (DRAM) technology. When performing computing, data in main memory needs to be processed. Because the CPU can only process the data in the cache, memory controller will send a series of instructions to the DRAM module through the off-chip bus, receive the response from the DRAM module after receiving the instructions, and then use the cache or registers to save the data \cite{mutlu2019processing}.

\subsubsection{Limitations of conventional computing}

Continuous expansion in the amount of computations and data leads to stricter requirements for high-performance, and computer architecture needs to shift from the intensive type of computing to the intensive type of memory \cite{tang2019bridging}. Nowadays, because of many limitations, it is hard for traditional computing to adapt to the current development pace.

\begin{figure}[hbt!]
\centering
\includegraphics[width=.75\textwidth]{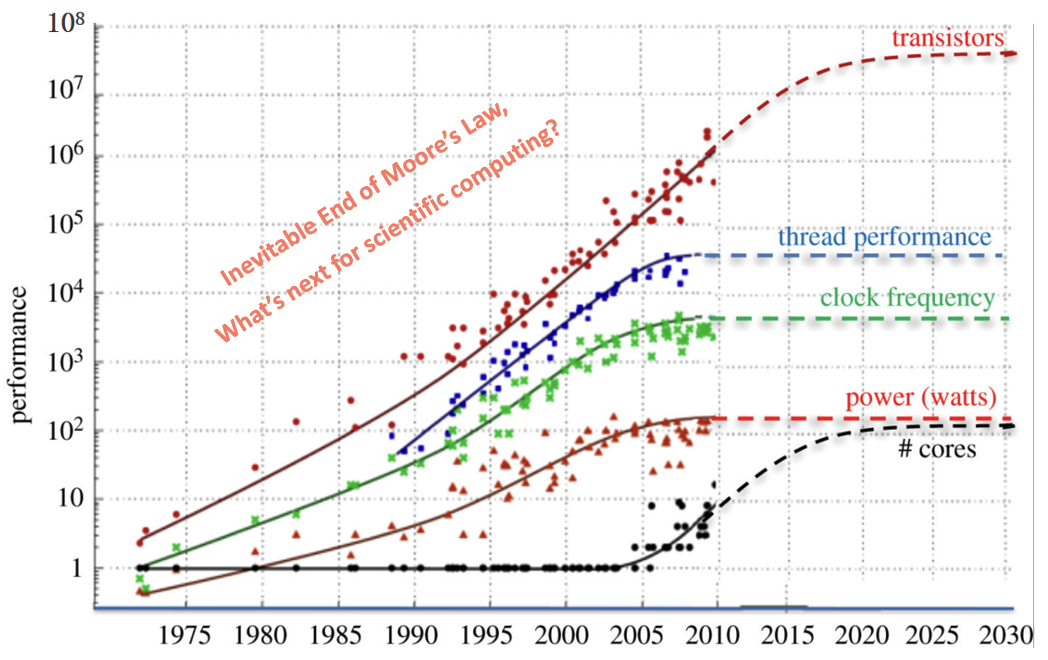}
\caption{Moore’s Law timeline, including Moore's Bend with transistors/CPU inflected with multi-Core CPUs beginning in 2005. Improvement of computing performance has been greatly challenged by the end of Dennard scaling in 2004. All additional approaches to further performance improvements will end in approximately 2025 due to the end of the roadmap for improvements to semiconductor lithography. Adapted with permission.\cite{olukotun2005future} Copyright 2005, ACM.}\label{fig:1}
\index{figures}
\end{figure}

On one hand, as the scaling degree of device feature size approaching its immutable physical limit, the trend of rapid growth in aspects of speed and integration of microelectronic devices represented by complementary metal oxide semiconductors (CMOS) has slown down \cite{tang2019bridging, cheng2021memory}. The 10-nm technology of Intel, originally scheduled to be launched in 2016, was delayed until 2019. Apple and Samsung, despite claiming to have reached 5-nm or even 3-nm in their nanofab,  can not keep up with Moore’s Law in updating their miniaturization technology \cite{leiserson2020there}. Obviously, the effects of improving traditional computational performance by reducing device size have gradually  weakened, and Moore’s Law that has influenced the growth of computational capability in electronic devices is inevitably coming to an end. In other words, the Post-Moore era is coming \cite{huang2020memory} (Figure 6).

On the other hand, the Von Neumann architecture on which traditional computing is built faces problems such as “ memory wall” that limits system performance, making it difficult to meet modern requirements for efficiency, energy consumption, density and cost in high-performance computing. Under the Von Neumann architecture, instructions and data are placed in the same memory, and instructions follow serial execution rules, so both of them can not be accessed simultaneously to avoid confusion in memory access \cite{nahmias2019photonic,yazdanpanah2013hybrid,cheng2021memory}. Meanwhile, the Von Neumann architecture features a separate memory and computing architecture, where memory unit and computing unit remain separate \cite{nahmias2019photonic}. During computing, data is frequently transferred between the memory unit and the processor unit, resulting in non-negligible delays and consuming a significant amount of energy. Up to now, under serial execution rules, data movement induces longer signal delay and increased energy losses, because of the huge gap between the operation speed of the CPU and the speed of accessing memory (the computing speed of the CPU has far exceeded the speed of accessing memory), the restriction caused by the bandwidth of the memory hierarchies, and the heat dissipation issue caused by unresolved leakage. These conditions ultimately result in insufficient utilization of hardware resources, increased energy consumption and decreased computational efficiency. For example, the AI facial recognition network developed by Google utilized a total of 16000 CPU cores on a three-day training session while consuming 100 kilowatts of power \cite{shaafiee2017overcoming,wang2022optical,tang2019bridging,yazdanpanah2013hybrid,cheng2021memory,mutlu2019processing,huang2020memory}. To alleviate these burdens, computing capability can be enhanced by separately increasing the bandwidth of the memory and using graphics processing unit (GPU) or AI accelerator. However, this approach has a limited benefit in terms of improving computing speed and energy consumption, and is not a long-term sustainable plan \cite{huang2020memory}. In addition, application-specific integrated circuits (ASIC) also play a role in addressing relevant issues. Compared with GPU, ASIC can significantly reduce energy consumption, but most of the consumption during operation is still wasted in the data movement rather than logic operation \cite{hamerly2019large}.

Overall, considering the ending of Moore’s Law and the limitations of the von Neumann architecture on the further development of modern computing facilities, there is a vital need to break through the core architectural bottleneck and seek alternative architectures and paradigms to build non-von Neumann systems to prominently strengthen computing performance \cite{shaafiee2017overcoming,neagu2023architectural}.

\subsection{Introduction to neuromorphic computing and its advantages}

\begin{figure}[hbt!]
\centering
\includegraphics[width=.9\textwidth]{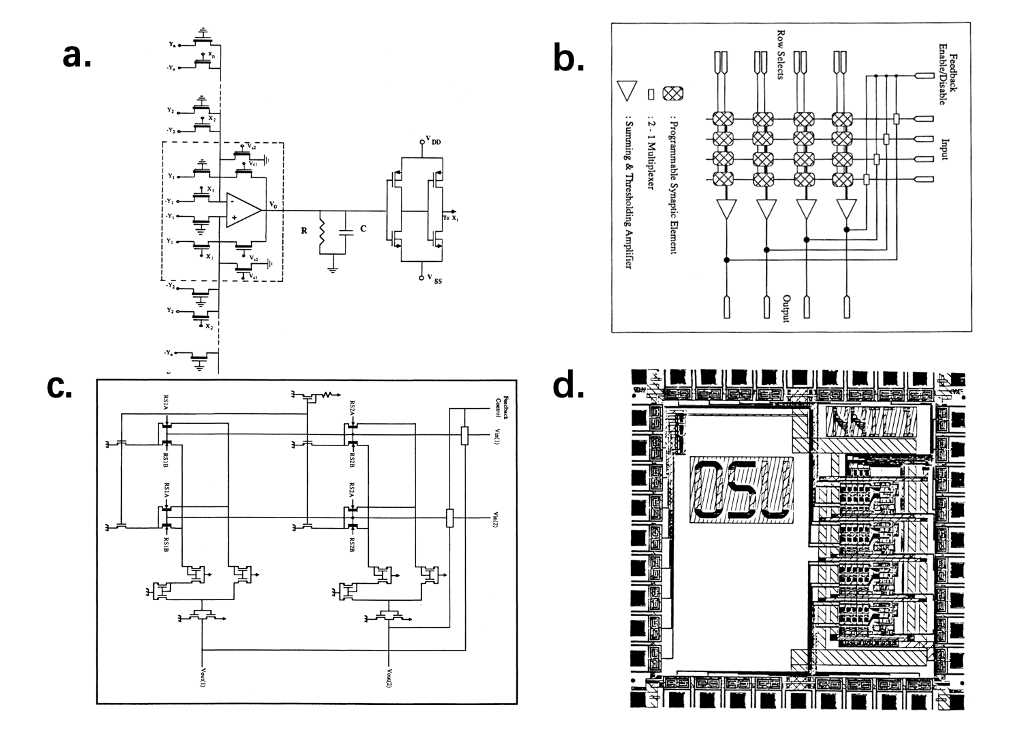}
\caption{Analog VLSI implementation of neural systems developed on or before 1989, marking the very first attempts to construct neuromorphic chips from electronics. a. A MOS circuit for the VLSI Implementation of Hopfield-like neural networks; b. Floating gate neural network; c. Two neuron circuit with four weight synapses; d. Layout of 4 neuron chip. Reproduced with permission.\cite{mead1989analog} Copyright 1989, Springer.}\label{fig:1}
\index{figures}
\end{figure}

Current computer technology is facing two important bottlenecks: the memory wall effect of the "von Neumann" architecture causes low energy efficiency \cite{wulf1995hitting,horowitz20141,ielmini2018memory}, and Moore's Law, which leads the development of semiconductors, is expected to expire in the next few years \cite{chirkov2023seizing,7878935,khan2018science,KISH2002144}. On one hand, the traditional processor architecture converts the processing of high-dimensional information into a one-dimensional processing of pure time dimension, which has low efficiency and high energy consumption \cite{zanotti2020smart}. This architecture cannot construct appropriate algorithms when processing unstructured information, especially when processing intelligent problems in real time.  In addition, the information processing takes place in the physically separated CPU and memory. Programs and data are sequentially read from the memory into the CPU for processing, and then sent back to the memory. This process causes a large amount of energy consumption \cite{zanotti2020smart}. A mismatch between the rate at which programs or data are transferred back and forth and the rate at which the CPU processes information results in a severe memory wall effect \cite{mckee2004reflections,saulsbury1996missing,wulf1995hitting}. On the other hand, as the semiconductor industry enters the sub-10 nm threshold, devices are approaching the limits of their physical shrinkage, and quantum effects are increasingly interfering with the normal operation of electronic devices \cite{goodnick2003quantum,gopi2022identification}. Although people have different estimates of the specific end time of Moore's Law, there is no controversy in the industry about the end of Moore's Law that has lasted for the past 50 years.

In 1989, Caltech's Carver Mead proposed the concept of "neuromorphic engineering (or brain-like computing)" in his book titled "Analog VLSI implementation of neural systems", which uses sub-threshold analog circuits to simulate Spiking Neural Network (SNN) \cite{mead1989analog}. As shown in Figure 7, Mead showed how the powerful organizing principles of animals' nervous systems can be realized in silicon integrated circuits, where examples include silicon neural systems that replicate animal senses. Meanwhile, Moore's Law continued to develop, and the frequency and performance of processors based on the von Neumann architecture continued to improve, while brain-inspired computing remained stagnant for more than 10 years. Around 2004, the main frequency of single-core processors stopped growing, and while the industry turned to multi-core processors,  the academic community began to seek alternative technologies such as non-von Neumann architectures. Since that point and in the following 10 years or so, neuromorphic computing has begun to attract widespread attention.

\begin{figure}[h]
\centering
\includegraphics[width=.9\textwidth]{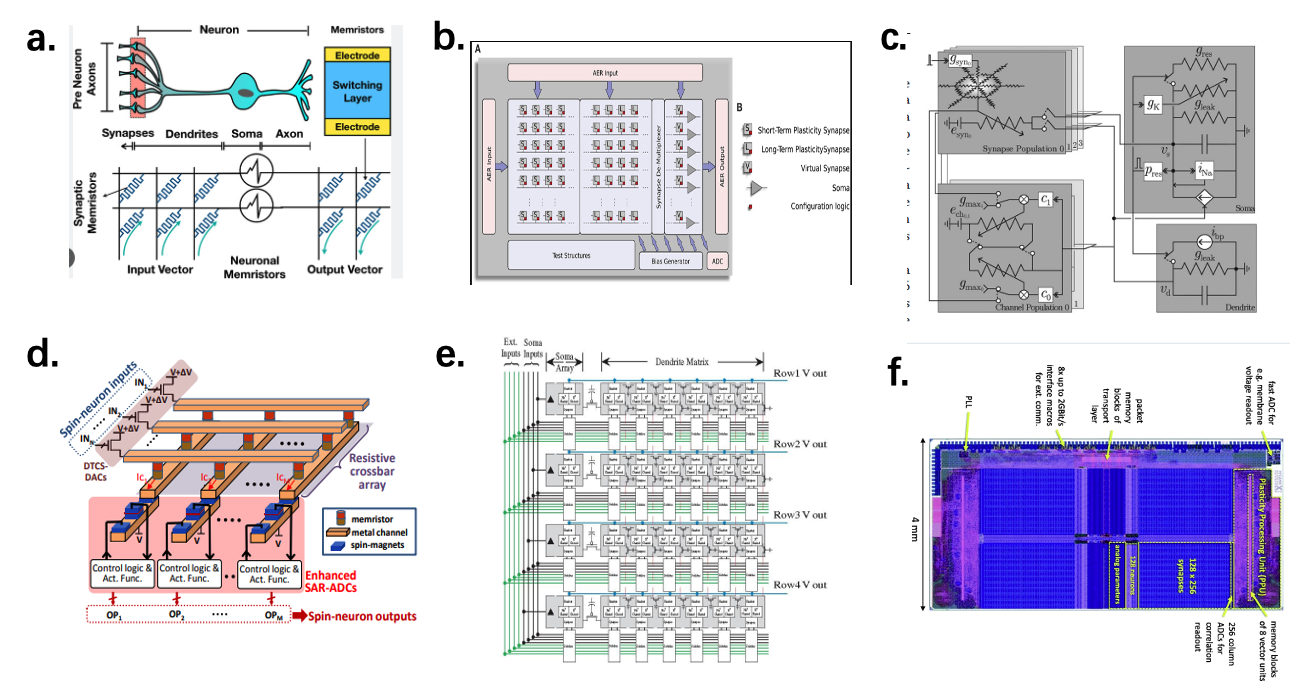}
\caption{Conventional neuromorphic computing systems, utilizing electronics rather than photonics. a. Neuromorphic Computing with Memristor Crossbar \cite{zhang2018neuromorphic}; b. A reconfigurable on-line learning spiking neuromorphic processor comprising 256 neurons and 128K synapses  \cite{qiao2015reconfigurable}. Block diagram of the architecture, showing two distinct synapse arrays (short-term plasticity and long-term plasticity synapses), an additional row of synapses (virtual synapses) and a row of neurons (somas); c. Neurogrid: A mixed-analog-digital multi-chip system for large-scale neural simulations \cite{benjamin2014neurogrid}; d. SPINDLE: SPINtronic deep learning engine for large-scale neuromorphic computing \cite{ramasubramanian2014spindle}. Shows an array of M spin-neurons that take N inputs each. The spin-neuron array consists of (i) Deep Triode Current Source Digital to Analog Converters that convert the N digital inputs into analog currents, (ii) a resistive crossbar array (N x M) that is used to perform weighted summation of the neuron inputs, and (iii) enhanced Successive Approximation Register ADCs that evaluate the activation function and produce the M digital outputs; e. A field programmable neural array (FPNA) \cite{farquhar2006field}. Each node of the matrix is identical, with the exception of the triangle wave generator on the output of the soma. Each node has some circuitry for readout, 1 Na+ channel, 1 K+ channel, 1 inhibitory synapse, and 1 excitatory synapse. Each node of the dendrite matrix is connected to its nearest neighbor in 2 dimensions; f. Layout of the current BrainScaleS-2 full-size ASIC \cite{grubl2020verification}. It contains 512 neuron circuits and 131072 synapse circuits which are arranged in 4 quadrants. Data lines of the synapse arrays and the column ADCs are directly connected to the PPUs at the top and bottom edges. Each PPU contains 8 vector units with dedicated memory blocks in addition to the general-purpose processor part. Panels reproduced with permission from: a. Reproduced with permission.\cite{zhang2018neuromorphic} Copyright 2018, Wiley. b. Reproduced under the terms of the Creative Commons Attribution License (CC BY) (http://creativecommons.org/licenses/by/4.0/).\cite{qiao2015reconfigurable} Copyright 2015, the Authors. c. Reproduced with permission.\cite{benjamin2014neurogrid} Copyright 2014, IEEE. d. Reproduced with permission.\cite{ramasubramanian2014spindle} Copyright 2014, ACM. e. Reproduced with permission.\cite{farquhar2006field} Copyright 2006, IEEE. f. Reproduced under the terms of the Creative Commons Attribution 4.0 International License (http://creativecommons.org/licenses/by/4.0/).\cite{qiao2015reconfigurable} Copyright 2020, the Authors and Springer. }\label{fig:1}
\index{figures}
\end{figure}

It is well known that the nervous system of mammals, especially humans, is one of the most efficient and robust structures in nature. The human brain has a large number of connections and exhibits strong parallelism. It has about $10^{11}$ neurons and $10^{15}$ synapses, but consumes only about 20W of energy \cite{Kuzum_2013}. Neurons achieve biological interconnection at a speed of a few milliseconds and have excellent fault tolerance mechanisms for component-level failures \cite{sherrington2023integrative}. For computer scientists, there are tremendous similarities between neural systems and digital systems. Components such as cell body, dendrites, axons, nerve terminals, and synapses together constitute a neuron unit. Specifically, the core part of the neuron is a cell body containing a nucleus, with a radius of 2 to 60 microns; there are two types of cell processes of different lengths on the surface of the cell body, which are long axons (only one) and short dendrites (usually multiple); the excitatory transmission between neurons passes through axons and nerve terminals and finally reaches the synapse (the connection point between neurons) \cite{kandel2000principles}. Neurons with various functions constitute a complete nervous system, which can effectively receive, integrate and transmit information/signal. This is considered to be the core link in the process of nervous system learning and adaptation. Although brain neural networks have different information processing and logical analysis capabilities at different levels, they are a coordinated and unified whole and are closely connected with each other \cite{kandel2000principles}. The neuromorphic computer is a novel computer model that simulates the operation of the brain's neural network with ultra large-scale pulses and real-time communication \cite{mead1990neuromorphic}. Neuromorphic computers simulate the high performance, low power consumption, real-time and other characteristics of biological brain neural networks, and use large-scale CPU/GPU clusters to implement neural networks. In the CPU cluster, each thread will map and simulate the corresponding neuron, and thousands of threads (neurons) run in an orderly coordinated manner to form a complete large-scale neural network.

The implementation of neuromorphic computing at the hardware level can be achieved through oxide-based (CMOS) devices such as memristors, spin electronic memories, threshold switches, transistors, etc. Some examples are shown in Figure 8. Back in 2006, researchers at Georgia Tech proposed a field-programmable neural array \cite{farquhar2006field}. The chip is the first in a series of increasingly complex floating-gate transistor arrays that allow the charge on the MOSFET gate to be programmed to simulate the channel ion properties of neurons in the brain, and is the first silicon programmable Neuron array. At the same time, many transistor-based brain-inspired chips and brain-inspired computer systems have also developed to a certain extent. For example, Stanford University developed the brain-inspired chip "Neurogrid" based on analog circuits, the University of Manchester began to develop SpiNNaker, a multi-core supercomputer that supports spiking neural networks based on ARM, the European Union's FACETS (fast analog computing with emergent transient states) project, and the U.S. Defense Research Agency DARPA's SyNAPSE (systems of neuromorphic adaptive plastic scalable electronics) project \cite{farquhar2006field,benjamin2014neurogrid,furber2014spinnaker,meier2004fast,hylton2008systems}. In 2008, HP realized a memristor prototype that could simulate the function of synapses, and demonstrated the first hybrid circuit of memristor and silicon materials \cite{xia2009memristor}. The global craze for artificial synapses began to take off in June 2012, when spintronics researchers at Purdue University published a paper on designing neuromorphic chips using lateral spin valves and memristors \cite{ramasubramanian2014spindle}. They believe the architecture works similarly to neurons and could therefore be used to test methods of reproducing brain processing. In addition, these chips are significantly more energy-efficient than conventional chips. In the same year Dr. Thomas and his colleagues at the University of Bielefeld created a memristor with learning capabilities. And in the following years \cite{thomas2015tunnel}, Thomas used this memristor as a key component of an artificial brain. Because of this similarity between memristors and synapses, it is an excellent material for building artificial brains—and thus a new generation of computers. Memristor allows us to build extremely energy-efficient, durable, and self-learning processors. It is precisely because of this ability that memristors have also received increasing attention in areas like neural networks and artificial intelligence. Several research groups are now exploring the use of memristors to build more efficient neural network architectures.

\subsection{Challenges and limitations faced by conventional neuromorphic computing }

The limitations of traditional brain-inspired computing mainly encompass three aspects: hardware limitations, software and algorithm limitations, and practicality and application limitations. They will be discussed in detail below in the text as well as in Figure 9, which compares biological brains, electronic, and photonic neuromorphic computing systems.

\begin{figure}[hbt!]
\centering
\includegraphics[width=.85\textwidth]{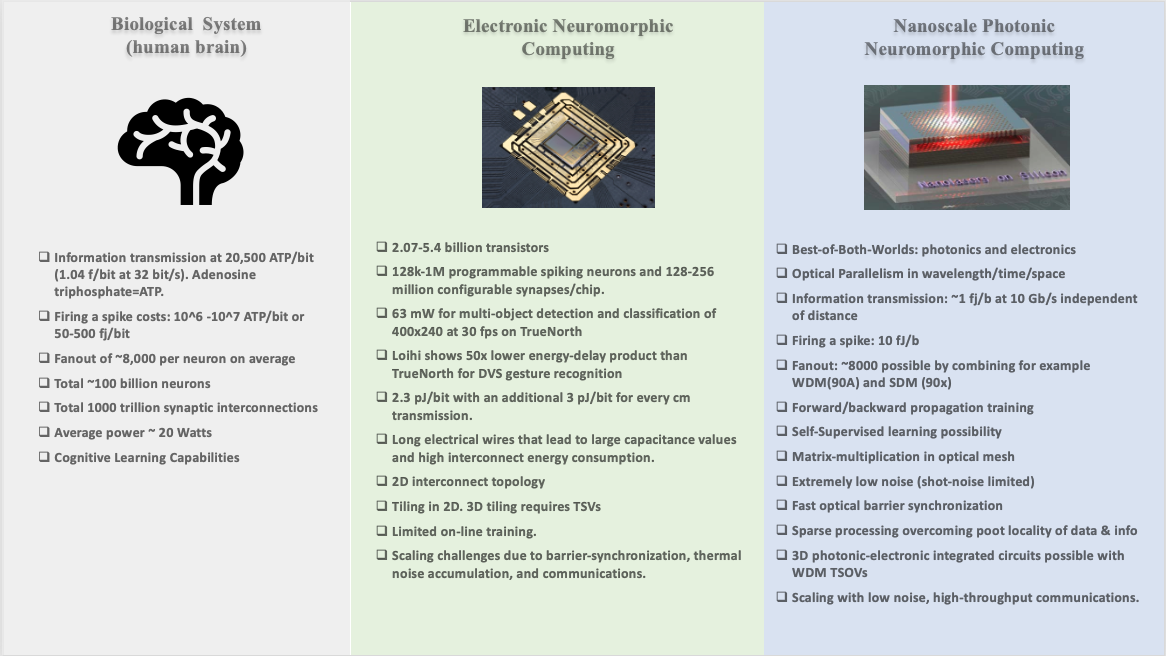}
\caption{Comparisons of a biological cognitive system (human’s cortex and brain, Figure 2a), CMOS-based electronic neuromorphic computing (E.g., IBM TrueNorth, Intel Loihi, Figure 8), and the latest photonic neuromorphic computing (Figure 3 and Table 1). Data in this figure are extracted from: \cite{henning2000spec,davies2018loihi,xiang2018numerical,laughlin2003communication,painkras2013spinnaker}. Reproduced under the terms of the Creative Commons Attribution License (CC BY) (http://creativecommons.org/licenses/by/4.0/).\cite{el2022photonic} Copyright 2022, the Authors and AIP Publishing.}\label{fig:1}
\index{figures} 
\end{figure}

Although neuromorphic computing advertises low power consumption, existing hardware typically consumes more energy than biological nervous systems. Decades of research and billions of dollars have been invested in various forms of pattern recognition, and while substantial improvements have been made, synthetic electronic systems still fall short of the capabilities of human perception on specific problems \cite{borji2014human,gelly2012grand}.  Materials play a key role in the energy consumption of neuromorphic computing. Due to the conductivity and resistivity characteristics of conventional conductors, a large amount of energy loss will inevitably occur during the transmission of electrical currents. Moreover, the size factor is also limiting the energy consumption limit of traditional conductive materials. As neuromorphic computing devices continue to shrink in size, quantum and thermal effects may become more pronounced, affecting energy efficiency \cite{li2021challenges,torres2023thermal}. As seen in Figure 9, conventional electronic systems typically consume three orders of magnitude more energy than biological systems and photonic systems.

In computing systems and networks, "Communication Overhead" is a term used to describe the additional resources and time required for information transmission \cite{macmillan2004communication}. It is an important factor affecting the computing performance and energy efficiency. For neuromorphic computing, this concept has several important aspects. First, data transmission is a core issue because simulating neurons in neural networks requires information transfer through synapses, which usually consumes additional energy and time in hardware implementation \cite{markovic2020physics}. Second, the activity of neurons usually needs to be represented through specific encoding and decoding methods, which also increases computational and energy overhead. In addition, some neuromorphic computing models require precise timing synchronization between neurons, which further increases communication and computing overhead \cite{park2020t2fsnn,schuman2020low}. In large-scale neural networks, information also requires complex routing between multiple neurons and network layers, which often requires additional hardware and algorithm support. In order to ensure the accuracy of information transmission, error detection and correction mechanisms need to be introduced, which will also bring more communication overhead \cite{hu2014memristor}. Finally, under high load or high concurrency, information may need to be queued in a buffer waiting to be processed, resulting in latency and additional energy consumption \cite{lu2020low,lee2022improved}. Therefore, communication overhead is an important factor affecting the performance and energy efficiency of brain-inspired computing, and its optimization usually requires comprehensive consideration from multiple aspects such as hardware design, network topology, and signal encoding methods. In Figure 9, it's shown that photonic systems can transmit data at a high rate and a low energy level independent of distance, which essentially solves the communication overhead of electronic systems.

Parallelism and synchronization also influence power consumption of neuromorphic computing. In traditional electronics, parallelism and synchronization face several major limitations. First, in terms of parallelism, electronic devices are often limited by circuit bandwidth, which affects their ability to perform highly parallel processing \cite{yan2022distributed,chen2016parallel}. Additionally, in highly parallel electronic systems, electromagnetic interference can become a serious problem, limiting overall performance \cite{painkras2013spinnaker}. Second, in terms of synchronization, maintaining global clock synchronization can be very complex and energy-intensive in large-scale electronic systems \cite{liu2021low}. In addition, the propagation speed of electronic signals in wires is limited, which further affects the accuracy of system synchronization. Therefore, these factors together constitute the main limitations of traditional electronic devices in terms of parallelism and synchronization. Photonic systems, on the other hand, can achieve high-degree of parallelism using the inherent fanout and WDM techniques , depicted in Figure 9.

In addition to the limitations of energy consumption of conventional neuromorphic computing, imitating large-scale neural networks requires a large amount of hardware resources, which is impractical in terms of space and cost.

To mimic large-scale biological neural networks, large numbers of neurons and synapses are needed. This results in larger hardware size or footprint, which increases cost and space requirements. For example, BrainScaleS is a nuromorphic computing system for large-scale simulations, which requires a full-sized dedicated computer room to accommodate it \cite{grubl2020verification}. Neurogrid, A hardware platform that simulates cortical neural networks, comes in a size equivalent to a standard 19-inch rack \cite{benjamin2014neurogrid}. On the other hand, quantum effects make it impossible to further reduce the size of transistors and therefore the size of brain-like computers \cite{gopi2022identification,zwanenburg2013silicon}. Moreover, the complex interconnection structure between neurons faces huge challenges in hardware implementation, especially when pursuing low power consumption and high performance. Specifically, neuromorphic computing involves a series of complex dynamics and nonlinear operations, which undoubtedly increases the complexity of hardware and algorithm design. At the same time, simulating biological neural networks usually requires storing a large amount of parameters and status information, which not only places higher requirements on storage resources, but also has a considerable impact on system size and energy consumption \cite{grubl2020verification}. As the network scale expands, how to effectively expand hardware scaling and maintain low energy consumption becomes particularly critical. In this context, integrating multiple components such as neurons, synapses, and learning rules into an efficient operating system becomes a difficult task. Finally, it is worth noting that existing semiconductor processing technology also has certain limitations in terms of integration density, power consumption and reliability \cite{zhang2018materials}. Therefore, solving these problems of size and complexity requires interdisciplinary research collaboration and continued technological innovation in the industry.

Last but not least, in neuromorphic computing, software and algorithms face a series of challenges and limitations as well. First, the software implementing these models is often extremely complex and computationally intensive due to the complex neurodynamics and nonlinear calculations involved \cite{schuman2022opportunities}. This complexity not only limits the feasibility of the algorithm in practical applications, but also increases the computational burden. Secondly, especially in applications based on spiking neural networks (SNNs), the algorithm training process often requires a large amount of time and computing resources, which poses an obvious obstacle to scenarios that require real-time or near-real-time response \cite{park2020t2fsnn}. Furthermore, compared with traditional machine learning algorithms, neuromorphic algorithms are often more difficult to interpret and verify, which is particularly problematic in applications that are highly sensitive and require interpretability, such as medical or autonomous driving \cite{ponulak2011introduction}. In addition, most of the existing brain-inspired algorithms are developed on traditional computing platforms that do not fully match the brain-inspired hardware architecture. This mismatch often leads to algorithm performance degradation in real hardware environments. At the same time, although the learning rules and plasticity mechanisms in biological neural networks are extremely complex, current algorithms fail to fully simulate these advanced characteristics, thus limiting their effectiveness in handling more complex real-world tasks \cite{mead1990neuromorphic}. Finally, although brain-like algorithms perform well on certain specific tasks, they usually lack the versatility and adaptability to demonstrate across multiple tasks and different environments.

\section{Overview of the Fundamentals of Photonics}

\subsection{Basics of photonics and light-matter interactions}

Photonics is a multidisciplinary field that studies light propagation and light–matter interactions, which has applications in various fields such as optical sensing \cite{Altug_Oh_Maier_Homola_2022}, optical interconnection, and optical communication \cite{Yao_Zheng_2023}. Photonics also plays an important role in developing optical computing systems, providing an alternative to electronic computers for high-speed and low-power data processing and transmission \cite{Wang_Ma_Wright_Onodera_Richard_McMahon_2022,Xingyuan_Xu_Xu_Tan_Corcoran_Wu_Boes_Nguyen_Chu_Little_Hicks_et}. This section introduces the basics of photonics in light-matter interactions.

Maxwell's equations constitute the cornerstone of classical electrodynamics, delineating the fundamental principles underlying the dynamical behavior of electric and magnetic fields. These equations encapsulate the generation of electromagnetic fields by charges and currents (Gauss's law for electricity and Ampère's law with Maxwell's addition), the non-existence of magnetic monopoles (Gauss's law for magnetism), and the induction of electric fields by time-varying magnetic fields (Faraday's law of induction). In the context of light-matter interactions, Maxwell's equations are indispensable. They predict the self-propagating nature of electromagnetic waves in a vacuum, a phenomenon in which light is a principal exemplar. Upon interaction with matter, these equations govern the complex processes of reflection, refraction, absorption, and transmission. The boundary conditions dictated by Maxwell's equations at interfaces between different media determine the electromagnetic field distributions and, consequently, the optical responses of materials. Furthermore, Maxwell's equations are integral to the quantification of the dielectric and magnetic properties of materials, characterized by permittivity and permeability, respectively. These properties influence the phase velocity of electromagnetic waves in media, leading to phenomena such as dispersion and polarization. The equations also describe the conservation of energy and momentum in electromagnetic fields, providing a framework for understanding the interaction forces and torques exerted by light on material particles, which is the basis for optical trapping and manipulation technologies.

Equations \eqref{eq:Maxwell1}--\eqref{eq:Maxwell4} are Maxwell's equations that dictate the behavior electromagnetic waves: \par
\noindent\begin{minipage}{.5\linewidth}
\begin{align}
    \bm{\nabla}\cdot\bm{D} &=4\pi\rho_{f}\;\label{eq:Maxwell1}
    \vphantom{\frac{\partial\bm{B}}{\partial t}}\\
    \bm{\nabla}\cdot\bm{B} &=0\;\vphantom{\frac{\partial\bm{B}}{\partial t}}\label{eq:Maxwell2}
\end{align}

\end{minipage}%
\begin{minipage}{.5\linewidth}
\begin{align}
    \bm{\nabla}\times\bm{E} &=-\frac{\partial\bm{B}}{\partial t}\;\label{eq:Maxwell3}
    \\
    \bm{\nabla}\times\bm{H} &=\bm{J}_f
    +\frac{\partial\bm{D}}{\partial t}\;\label{eq:Maxwell4}
\end{align}
\end{minipage}
\smallskip

where D is electric flux density, B is magnetic flux density, E is electric field intensity, and H is magnetic field intensity. Maxwell's equations form the foundation of classical photonics and light-matter interactions.  

To build efficient optical neural networks (ONN), nanoscale components are required to manipulate and process light signals. Diffraction and interference of light at the nanoscale indeed play a crucial role in these components. These phenomena enable precise control over light signals, allowing for information encoding, processing, and transmission within the network.
In principle, diffraction and interference both arise due to the superposition principle, which states that when two or more waves overlap, the resulting wave is the sum of their individual amplitudes. For example, the photonic oscillations $E_1$ and $E_2$ produced at a certain point by coherent light waves can be expressed as,

$$
E_1=a_1\ \exp{\left[i\left(\alpha_1-\omega t\right)\right]} 
$$ 
$$
E_2=a_2\ \exp{\left[i\left(\alpha_2-\omega t\right)\right]}
$$

So, it can be readily deduced that the resultant superimposed oscillation is 

$$
E_s=a\exp{\left[i\left(\alpha-\omega t\right)\right]}
$$
Where 
$$
a^2=a_1^2+a_2^2+2a_1a_2\cos{\left(\alpha_2-\alpha_1\right)}
$$
$$
tan{\alpha}=\frac{a_1sin{\alpha_1}+a_2sin{\alpha_2}}{a_1cos{\alpha_1}+a_2cos{\alpha_2}}
$$

It is obvious that by varying $\alpha_1$ and $\alpha_2$, both the amplitude \(a\) and phase $\alpha$ of the combined light wave are changed. By architecting specialized dielectric material structures on a nanoscale, intricate photonic interactions can be actualized.

For example, B. Wu et. al. actualized a variety of nonlinear activation functions through the employment of the thermo-optic effect and micro-ring resonators \cite{Bo_Wu_Hengkang_Li_Weiyu}. They use germanium whose absorption coefficient $\alpha$ and refractive index \(n\) are characterized by temperature \(T\) as follows:

$$
\Delta\alpha=k_1\exp{\left(\Delta T\right)}
$$
$$
\Delta n=k_2\Delta T
$$

Where $k_1$ and $k_2$ are constant and $\Delta T$ is change of temperature. These two values have influence on quality factor of micro-ring resonators, and thus change the output power of it.
Another example is MZI, which is a device that utilizes the principle of light interference to measure small changes in phase or refractive index \cite{amin20180}. It consists of a beam splitter that splits an incoming light beam into two paths, which then recombine at a second beam splitter. However, we can also utilize the electro-optic effect and nonlinear effects to alter the refractive index and phase, modify the interference conditions in MZI, and finally realize the change of light wave amplitude. 

\subsection{Comparison between photons and electrons}
Electrons and photonics are both elementary constituent particles of the physical world. Electrons are fermions which are particles with half-integer
spins and they obey Pauli's exclusion principle. This means that no two fermions can occupy the same energy state.  Other fermions include: Protons, Neutrons, Neutrinos, Quarks.  Fermions are different from bosons, which have integer spins and do not obey Pauli's exclusion principle. Some examples of bosons include: photons, a-particles, Higgs, Helium atoms, and Gluons. More quantitatively, bosons are the fundamental particles that have spin in integer values (0, 1, 2, etc.). Fermions, on the other hand, have spin in odd half integer values (1/2, 3/2, and 5/2, but not 2/2 or 6/2).

Wave function of electrons is prescribed by the Schrodinger equation (Equation v), while that of photons is derived from the Maxwell's equations in section 3.1 and is sometimes referred to as the plane wave solutions (Equation vi-vii). Electrons have a specific mass and a drift velocity of limited magnitude, whereas photons are massless and travel at the speed of light.

The Schrödinger equation for the electron is:
\begin{align}
    E \psi = -\frac{\hbar^2}{2\mu}\nabla^2\psi - \frac{q^2}{4\pi\varepsilon_0 r}\psi 
\end{align}

where $E$ is energy, $q$ is the electron charge, $\mathbf{r}$ is the position of the electron relative to the nucleus, $r = |\mathbf{r}|$ is the magnitude of the relative position, the potential term is due to the Coulomb interaction, wherein $\varepsilon_0$ is the permittivity of free space and
\[\mu = \frac{m_q m_p}{m_q+m_p} \]
is the 2-body reduced mass of the hydrogen nucleus (just a proton) of mass $m_p$ and the electron of mass $m_q$. 

The planar traveling wave solutions of the photon wave equations are:
\begin{align}
\mathbf{E}(\mathbf{r}) &= \mathbf{E}_0 e^{ -i \mathbf{k} \cdot \mathbf{r} } \\
\mathbf{B}(\mathbf{r}) &= \mathbf{B}_0 e^{ -i \mathbf{k} \cdot \mathbf{r} }
\end{align}

where \(\mathbf{r} = (x, y, z)\) is the position vector (in meters) and $\mathbf{k}$ is the wavenumber.

In semiconductor physics, electrons and photons are closely intertwined.  A diode laser or a light-emitting diode (LED) is a semiconductor device that emits light when electrical current flows through it. As A. Einstein correctly predicted in 1916, electrons in the semiconductor can recombine with electron holes, releasing energy in the form of photons. The color of the emitted light (corresponding to the energy of the photons) is determined by the energy required for electrons to cross the band gap of the semiconductor.

\section{Photonic Components for Neuromorphic Computing}

\subsection{Key devices of photonics for neuromorphic computing} 

Nano-photonics, as an emerging interdisciplinary subject, integrates the principles of nanotechnology and photonics, aiming to explore and harness the manipulation of light wave by nanoscale structures. In the landscape of photonics, active devices and passive devices are crucial and have broad application prospects. Neuromorphic systems aim to mimic the computational and cognitive capabilities of the brain by leveraging the principles of neural networks. In this part, we will focus on the key devices in photonics, divided into two parts: active devices and passive devices, and provide an in-depth analysis of their working principles, applications in neuromorphic computing, and future development trends. Some representative devices are shown in Figure 10.

\begin{figure}[hbt!]
\centering
\includegraphics[width=.85\textwidth]{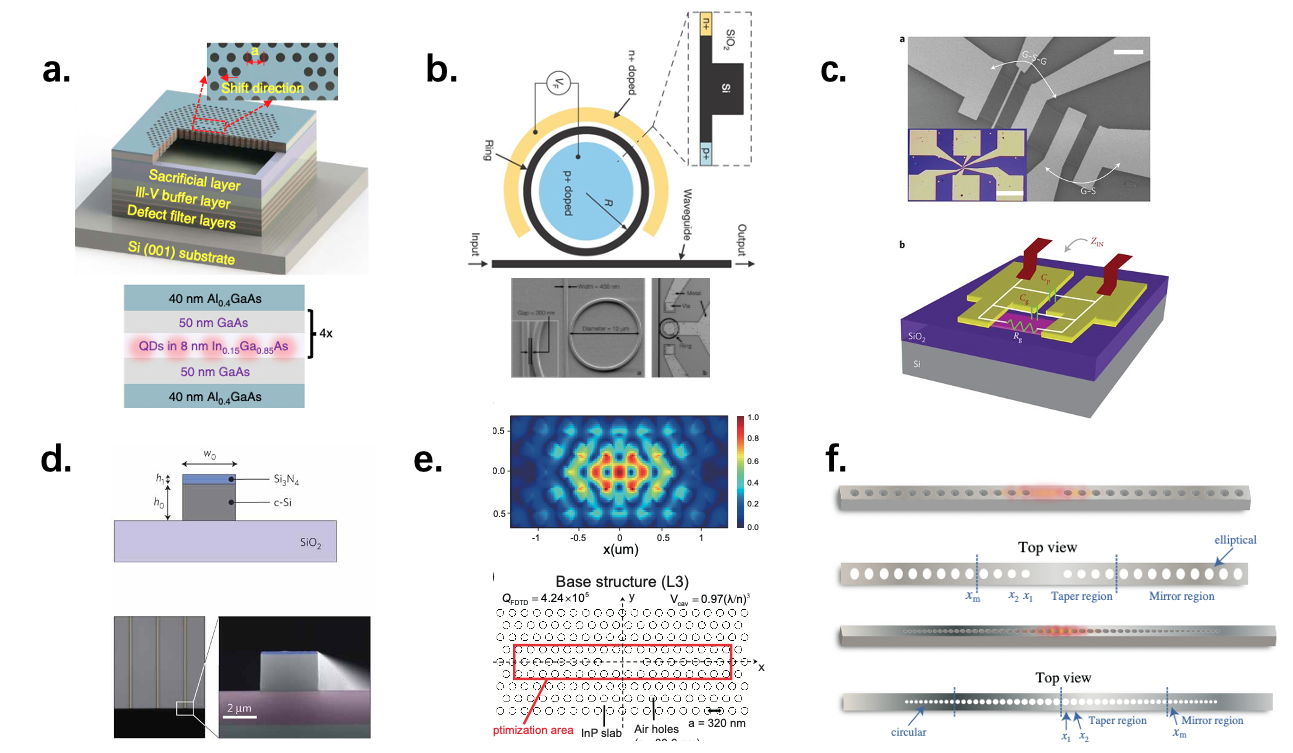}
\caption{Key photonic devices and structures for neuromorphic computing. a. Schematic diagram of the fabricated InAs/GaAs QD PC (L3 cavity) laser epitaxially grown on on-axis Si (001) substrate \cite{zhou2020continuous}. The lattice constant, radius and shift of L3 PC cavity are a, r and 0.15a, respectively. 
b. Schematic layout of the ring resonator-based modulator \cite{xu2005micrometre}. The inset shows the cross-section of the ring. R, radius of ring. VF, voltage applied on the modulator. SEM and microscope images of the fabricated device. 
c. SEM and optical (inset) images of the high-bandwidth graphene photodetectors \cite{xia2009ultrafast}. The graphene shown here has two to three layers. Two types of wirings are shown: ground–signal (G–S) and ground–signal–ground (G–S–G). The high-frequency results are from devices with G–S wirings. 
d. Schematic representation of the cross-section of a strained SOI waveguide \cite{cazzanelli2012second}. Si forms the core of the waveguide and the parameters are chosen to have more than 95\% of the optical field confined within the waveguide. 
e. The L3 photonic crystal resonance nanocavity \cite{li2021deep}. f. The nanobeam photonic crystal structure \cite{li2023deep}. Panels reproduced with permission from: a. Reproduced under the terms of the Creative Commons Attribution 4.0 International License (http://creativecommons.org/licenses/by/4.0/).\cite{zhou2020continuous} Copyright 2020, the Authors and Springer Nature. b. Reproduced with permission.\cite{xu2005micrometre} Copyright 2005, Springer Nature. c. Reproduced with permission.\cite{xia2009ultrafast} Copyright 2009, Springer Nature. d. Reproduced with permission.\cite{cazzanelli2012second} Copyright 2012, Springer Nature. e. Reproduced under the terms of the Creative Commons Attribution 4.0 International License (http://creativecommons.org/licenses/by/4.0/).\cite{li2021deep} Copyright 2021, the Authors and Optica. f. Reproduced under the terms of the Creative Commons Attribution 4.0 International License (http://creativecommons.org/licenses/by/4.0/).\cite{li2023deep} Copyright 2023, the Authors and De Gruyter. 
}\label{fig:1}
\index{figures}
\end{figure}

1. Active devices

Active devices are photonic devices that can generate, manipulate or amplify optical signals, including lasers, optical modulators, photodetectors, etc. \cite{iqbal2021nanophotonics} These devices play a vital role in fields such as communications, sensing, laser processing, and medical diagnosis.

1.1 Laser

Lasers are the most important and widely used active devices in photonics and are known for their highly coherent light output, high emission power, and high spectral brightness \cite{zhou2020continuous}. In the current state of photonics, different types of lasers such as Vertical Cavity Surface Emitting Lasers (VCSELs), DFB lasers, micro-ring lasers, quantum dot lasers, and the latest topological lasers and PCSELs (see section 6.1 for details) \cite{li:hal-04175312,hirose2014watt} are developing rapidly \cite{kneissl2020semiconductor}. These lasers not only play a key role in optical communications, but are also used in fields such as Lidar, optical sensing and medical imaging. Lasers can be pulsed, continuous-wave, optically or electrically pumped, depending on their specific needs. 

1.2 Optical modulator

Light modulators are used to control the intensity, phase and frequency of light in real time \cite{xu2005micrometre}. In photonics, technologies such as electro-optic modulators and acousto-optic modulators are widely used in high-speed optical communication systems to achieve high-speed data transmission and information processing. Nanoscale optical modulators offer the advantages of small size, low power consumption and high speed and are key components of next-generation communication systems and photonic computers \cite{ohtsu2002nanophotonics}.

1.3 Light detector

Photodetectors are important active devices of photonics, used to convert optical signals into electrical signals \cite{xia2009ultrafast}. There is a wide variety of photodetectors available, including photodiodes, photodetection arrays, and single-photon detectors. These devices play pivotal roles in communications, sensing, optical imaging, and quantum information processing for next-gen photonic computing systems \cite{prasad2004nanophotonics,monticone2017metamaterial}.

1.4 Future Prospects of Active Devices

In the future, active devices will continue to grow and expand in photonics. The miniaturization, low power consumption and high efficiency of lasers will empower more applications, including quantum information processing and PICs. Optical modulators will continue to play a critical role in high-speed communications and Lidar. Photodetectors will usher in higher sensitivity, faster response times, and wider wavelength ranges, driving innovation in fields such as wireless communications, medical diagnosis, and optical sensing. All these devices will substantially contribute to the growth of neuromorphic computers in the near future.

2. Passive components

Passive devices refer to photonic devices that cannot generate or amplify optical signals, but they are equally essential for the transmission, control and processing of light. Common passive devices include waveguides, resonators, photonic crystals, etc.

2.1 Waveguide

A waveguide is a structure used to guide light in a desired direction, usually consisting of a high-refractive index material surrounded by a low-refractive index cladding \cite{cazzanelli2012second}. There are many types of waveguides, including optical fiber waveguides, planar waveguides, photonic crystal waveguides, etc. They play a key role in optical communications, optical sensing, photonic chips and other fields \cite{karabchevsky2020chip}. The low transmission loss of fiber optic waveguides makes them ideal for long-distance communications, while planar waveguides are often used in optical chips to realize micro-nanoscale integrated optical components \cite{roco2011applications}.

2.2 Resonant cavity

A resonant cavity is a device used to enhance light modes of specific frequencies \cite{li2021deep}. There are many types of resonant cavities, including fiber cavities, micro-ring cavities, micro-beam cavities, etc. Resonators can be used in filtering, lasing and sensing applications, and their high Q-factors give them a special place in photonics. Micro-ring cavities achieve resonance through multiple photon reflections and are widely used in micro-lasers, sensors and quantum information processing \cite{xi2011nano,ohtsu2008nanophotonics,dai2016silicon}. They can also be used along with soliton microcombs to perform wavelength division multiplexing. A nanobeam cavity \cite{deotare2009high} is an elongated waveguide with a micron-scale cross-section that can be used to achieve a high-quality factor resonator whose size and shape can be tuned for specific applications. Key performance metrics of resonant cavities are the Q-factor and modal volume.

2.3 Photonic crystal

Photonic crystal is a material with a periodic structure that can realize the light band gap and waveguide mode of light \cite{Ma_Zhou_Tang_Li_Zhang_Xi_Martin_Baron_Liu_Zhang_et,xie2020higher}. It has a periodically varying refractive index enabled by an array of holes or rods. These band gap and waveguide modes can be used in optical filtering, lasers, sensors, and other applications \cite{alias2018review,momeni2009silicon}. Photonic crystal design and preparation technology plays an important role in laser designs, such as the popular Photonic Crystal Surface Emitting Lasers (PCSEL) \cite{hirose2014watt} and the more recent topological insulator lasers.

2.4 Future prospects of passive components

In the future, the development of passive devices will continue to promote the development of neuromorphic computing technologies. The design of the waveguide will be more flexible to meet the needs of different applications. The resonant cavity will further improve the quality factor and allow optical devices to operate more efficiently. The structure and function of photonic crystals will be further expanded to meet the requirements of a wider range of engineering applications. The continued development of these passive devices will bring more possibilities to the future of photonics-enabled areas.

We can agree that active devices such as lasers, optical modulators, and photodetectors provide optical signal generation and processing capabilities, supporting applications such as optical communications, medical imaging, and quantum information processing. Passive devices such as waveguides, resonators and photonic crystals provide key tools for light transmission and manipulation, supporting the development of optical communications, sensing applications and data transmission. In the future, these key devices will continue to be enhanced while photonics continues to drive cutting-edge research in neuromorphic computing, opening up new possibilities for future scientific and engineering applications.

\subsection{Optical interconnects for scalable neuromorphic systems}
Neuromorphic systems, inspired by the structure and function of the brain, have attracted significant attention in the AI and computational neuroscience community. Neuromorphic systems aim to mimic the computational and cognitive capabilities of the brain by leveraging the principles of neural networks. As the complexity and scale of these systems increase, the need for efficient, high-bandwidth connectivity becomes increasingly important, because providing efficient and scalable connections to support the massive flow of data between neurons and synapses is critical \cite{xu2022scalable}. Traditional electronic interconnects, despite their widespread use, still face bottlenecks in bandwidth, power consumption, crosstalk, and latency as neuromorphic systems evolve. Optical interconnects, which use photons to transmit information, have emerged as a promising solution to address the limitations of traditional electronic interconnects in neuromorphic computing. In this part, we provide an overview of recent advances in optical interconnects for scalable neural systems, highlighting key advances, challenges, and prospects. We draw insights from numerous research papers, focusing on advances in photonic components, network topologies, and system integration. Some representative optical interconnects are shown in Figure 11.

\begin{figure}[hbt!]
\centering
\includegraphics[width=.85\textwidth]{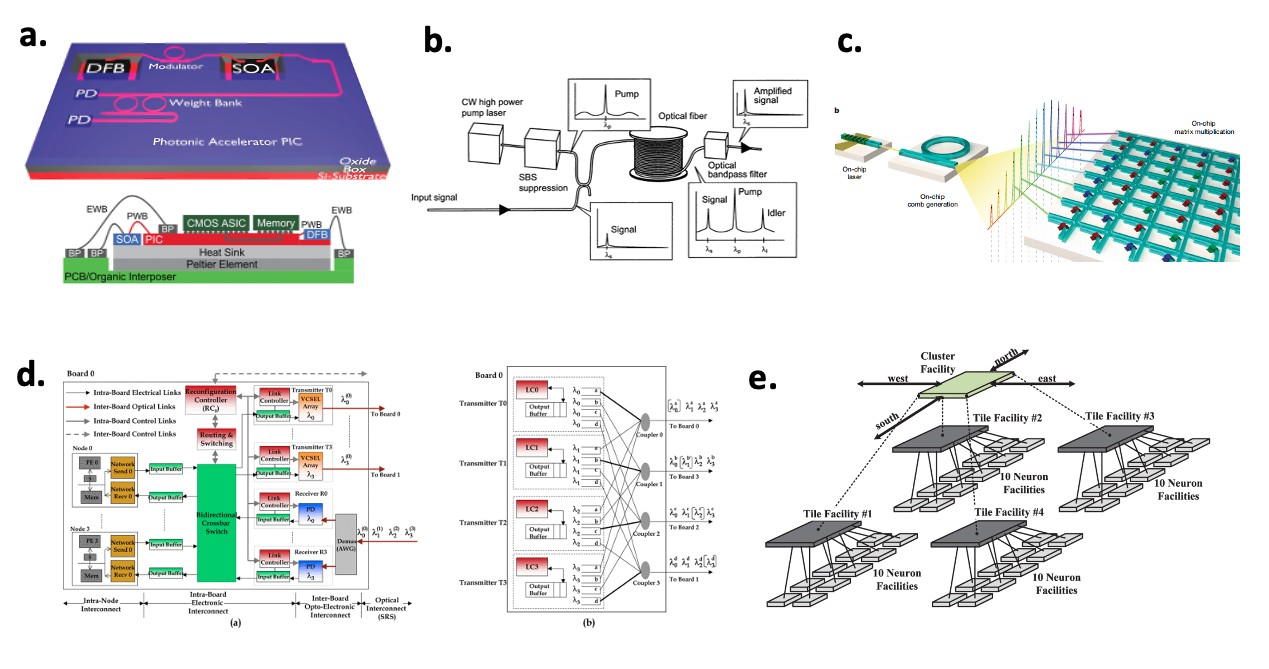}
\caption{Optical interconnects for scalable neuromorphic systems. a. Neuromorphic photonic accelerator PIC \cite{nezami2022packaging}. Prospect of a packaging solution consisting of a silicon PIC, CMOS chip, and III-V DFB lasers and SOAs integrated via photonic wire bonds and multi-chip integration. b. General scheme of phase-insensitive fiber-based optical parametric amplifier \cite{hansryd2002fiber}. c. Photonic tensor core \cite{feldmann2021parallel}. Conceptual illustration of a fully integrated photonic architecture to compute convolutional operations. An on-chip laser (not included here) pumps an integrated \ce{Si3N4} microresonator to generate a broadband soliton frequency comb. d. Proposed on-board interconnect for the E-RAPID architecture with RC and LCs and proposed technology for reconfiguration using passive couplers and array of lasers per transmitter port \cite{kodi2010energy}. e. Proposed scalable NoC architecture for implementing clusters of neurons to implement spiking neural networks \cite{carrillo2012scalable}. It aims to address the scalability issue by creating a modular array of clusters of neurons using a hierarchical structure of low and high-level routers. Panels reproduced with permission from: a. Reproduced with permission.\cite{nezami2022packaging} Copyright 2022, IEEE. b. Reproduced with permission.\cite{hansryd2002fiber} Copyright 2002, IEEE. c. Reproduced with permission.\cite{feldmann2021parallel} Copyright 2021, Springer Nature. d. Reproduced with permission.\cite{kodi2010energy} Copyright 2010, IEEE. e. Reproduced with permission.\cite{carrillo2012scalable} Copyright 2012, IEEE. 
}\label{fig:1}
\index{figures}
\end{figure}

\subsubsection{Advancements in Photonic Components}

The photonic components are fundamental components of optical interconnects. Recent innovations in this field have significantly improved the performance and efficiency of optical interconnects for neuromorphic systems. 

\textbf{Photonic integrated circuits (PIC)}\\
Photonic integrated circuits are compact and complex chips that combine multiple optical functions on a single chip. From the Shannon-Hartley theorem, PICs have an unlimited bandwidth $\approx$ 40THz which indicates a promising future of PICs to transport and process analogue signals. Recent developments in PICs have led to improved signal processing capabilities, allowing the implementation of complex neural network architectures. PICs also enable efficient signal routing, switching, and conditioning in neuromorphic systems \cite{nezami2022packaging,moralis2022coherent}. 

\textbf{Nonlinear Optical Devices}\\
Nonlinear optical devices, such as optical parametric amplifiers \cite{hansryd2002fiber} and inverters, have found application in improving the efficiency of optical interconnects. They enable wavelength conversion and amplification of optical signals, reducing the need for power-consuming electronic repeaters \cite{dabos2022neuromorphic}. 

\textbf{Photonic Neural Network Accelerators}\\
Research is underway to develop specialized photonic devices to accelerate neural network (deep learning) computations. Optical processors based on metasurface layers, soliton microcombs, VCSELs, and ring resonators can perform MVM at extremely high speeds, significantly improving the efficiency of deep learning in neuromorphic systems \cite{shastri2021photonics,markovic2020physics,feldmann2021parallel}. 

\subsubsection{Evolution of network topology }

Network topology in neuromorphic systems is essential for efficient learning and information transfer. Recent research has explored new optical network architectures to improve scalability and system performance. 
 
\textbf{Reconfigurable optical interconnects}\\
Reconfigurable optical interconnects provide flexibility in neural network configuration. By dynamically changing the connections between neurons and synapses, these connections enable rapid adaptation and learning \cite{nahmias2018neuromorphic,dangel2018polymer,kodi2010energy,xiao2019silicon}. Research has focused on developing tunable optical components and switches to facilitate reconfiguration. 

\textbf{Neural Network-on-Chip (NoC)}\\
Neural NoCs are specialized optical interconnects designed to efficiently connect various processing elements, including neurons and synapses. They reduce communication congestion and provide low-latency, high-bandwidth connections \cite{carrillo2012scalable}. Recent developments have demonstrated the benefits of NoC in large-scale nervous 
systems. 

\subsubsection{System Integration and Challenges}

The integration of optical interconnects into neuromorphic systems poses several challenges related to compatibility, scalability, and power consumption. 

\textbf{Hybrid Integration with Electronics}\\
Integrating optical interconnects with existing electronic neuromorphic systems is a complex task. Researchers are working on hybrid integration techniques to ensure seamless communication between electronic and photonic components, maximizing the benefits of both technologies \cite{zhang2020scalable,guo2021integrated}. Heterogeneous and monolithic integration of semiconductor materials in optical interconnects is a branch of technique that is alternative to hybrid integration. 

\textbf{Scalability}\\
Scalability is a key challenge for optical interconnects. As neuromorphic systems grow in size, ensuring that optical components can be manufactured and interconnected at scale without a significant increase in cost and complexity is essential \cite{shainline2017superconducting,catthoor2018very}. 

\textbf{Power consumption}\\
Although optical connections are more energy efficient than electronic connections, minimizing power consumption is still a top priority. Research is ongoing to develop low-power photonic components and advanced power management techniques \cite{liu2021low} for neuromorphic systems.

In short, optical interconnects have the potential to revolutionize neuromorphic computing by providing scalable, high-bandwidth, and energy-efficient communications paths. As research on photonic components, network topologies, and system integration continues to bloom, we can expect significant advances in the development of large-scale neuromorphic systems. Integrating optics into neuromorphic hardware promises to accelerate advances in AI, cognitive computing and a range of applications, including robotics, autonomous vehicles and biomedical research. Recent advances in these three areas discussed above are paving the way for efficient and scalable neuromorphic systems. While challenges regarding compatibility, scalability, and power consumption remain, the promise of optical interconnects to revolutionize neuromorphic computing is becoming increasingly clear.

\subsection{Optical logic gates in neuromorphic computing}

Similar to electronic computing, in optical computing, logic gate serves as the very bedrock upon which systems are constructed. In recent years, notable advancements have been achieved for optical logic gates, which have paved the way for enhanced neuromorphic computing \cite{Jiao_Liu_Zhang_Yu_Zuo_Zhang_Zhao_Lin_Shao_2022}.
The realization of optical logic gates have many different strategies, such as diffractive neural networks \cite{Lin_Rivenson_Yardimci_Veli_Luo_Jarrahi_Ozcan_2018}, semiconductor optical amplifiers \cite{Hu_Zhang_Zhao_2017,Jiao_Liu_Zhang_Yu_Zuo_Zhang_Zhao_Lin_Shao_2022,Kim_Kang_Kim_Han_2006} and photonic crystal waveguide \cite{li2023deep}.

Diffraction neural networks largely rely on the physical phenomenon of diffraction for computation. Typically, these networks undergo a pre-training process wherein they learn the intricate relationship between the input light field and the resulting patterns. This learning procedure is accomplished by adjusting the phase delay or transmission rate at each specific point in the network. Diffraction neural networks are generally composed of multiple metasurfaces \cite{Lin_Rivenson_Yardimci_Veli_Luo_Jarrahi_Ozcan_2018,Liu_Ma_Luo_Hong_Xiao_Zhang_Miao_Yu_Cheng_Li_et,Luo_Hu_Ou_Li_Lai_Liu_Cheng_Pan_Duan_2022}, each layer equating to a layer of neurons. With suitable training, manufacturing the corresponding diffractive network layers, and assembling them rightly, it can execute the required operations on any given input light field in real-time while performing neural network calculations at the speed of light. Diffraction neural networks operating in the mid-infrared are anticipated to function within a scale of a few millimeters. This endows it with immense prospects for large-scale integrated applications.

Semiconductor optical amplifiers (SOAs) are used to form optical logic gates due to their ability to modulate the intensity of light, change the phase of light, or both \cite{Hu_Zhang_Zhao_2017}. They can perform several different logical operations based on the principle of gain saturation, where the output light intensity is dependent on the input light intensity. This characteristic makes SOA highly valuable for building optical logic gates, enabling it to implement various applications such as all optical lattices, pseudo-random bit sequence generation, and all optical encryption.

Photonic crystal waveguides are structures that can limit and guide light in periodic dielectric media. They can control the propagation of light in a highly precise manner, making them an excellent platform for implementing optical logic gates. Especially in recent years, the emerging topological photonic crystal waveguides \cite{tan2021topological,Tang_He_Shi_Liu_Chen_Dong_2022,Wang_Sun_He_Tang_An_Wang_Du_Zhang_Yuan_He_et,Wang_Tang_He_Wang_Li_Sun_Zhang_Yuan_Dong_Su_2022} have significant advantages in the construction of all-optical logic gates due to their unidirectional propagation, controllable splitting, and other attractive characteristics.

These advancements in optical logic gates above have made optical neuromorphic computing a promising field for the development of high-speed, low-energy, and compact computing systems. In more recent years, researchers have begun to explore more sophisticated photonic devices such as frequency microcombs \cite{feldmann2021parallel}, micro-ring resonators \cite{xu202111}, phase change materials \cite{feldmann2021parallel}, nanolasers \cite{chen2023deep}, metasurfaces \cite{liu2022programmable}, optical attenuators \cite{ashtiani2022chip}, EDFA \cite{xu202111}, and photodiodes \cite{ashtiani2022chip} ] to achieve state-of-the-art neuromorphic computing systems. These will be discussed in detail in the remaining text.

\subsection{State-of-the-art fabrication platforms for photonic devices}

Nano-photonics is a multidisciplinary field concerned with manipulating light at the nanoscale, often on the order of the wavelength of light. This field has grown rapidly due to its potential applications in telecommunications, sensing, imaging, and computing. Photonic devices offer advantages such as higher speed, lower power consumption, and compact size, but making these devices requires advanced manufacturing techniques capable of the precise drilling and etching of fine nanostructures. 
In this part, we provide an overview of the state-of-the-art manufacturing platforms for photonic devices. We explore recent developments in lithography, etching, self-assembly and 3D printing, demonstrating their potential to revolutionize photonic technologies.

\begin{figure}[hbt!]
\centering
\includegraphics[width=.85\textwidth]{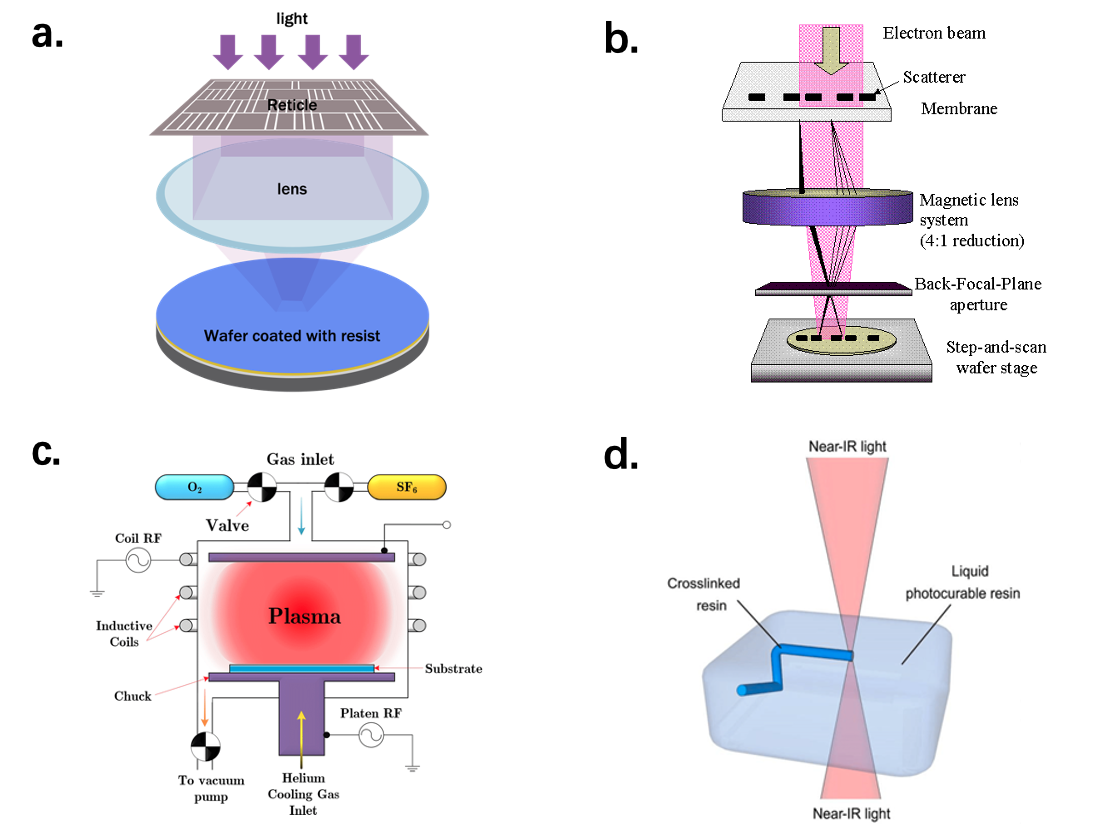}
\caption{State-of-the-art fabrication methods for photonic devices. a. Photolithography.
b. Electron beam lithography (EBL). c. ICP-RIE etching. d. 3D printing with two-photon polymerization (TPP). Panels reproduced with permission from: c. Reproduced with permission.\cite{plakhotnyuk2017low} Copyright 2017, AIP Publishing. d. Reproduced under the terms of the Creative Commons Attribution (CC BY) license (http://creativecommons.org/licenses/by/4.0/).\cite{yu20203d} Copyright 2020, the Authors and MDPI. 
}\label{fig:1}
\index{figures}
\end{figure}

\subsubsection{Lithography}

Lithography plays an important role in the production of optical devices and includes a variety of methods, including traditional photolithography, extreme ultraviolet (EUV) lithography, electron beam exposure and nano-printing technology.
    
First of all, traditional photolithography \cite{levenson1982improving} is a popular technology for manufacturing optical devices. It uses a UV light source and a mask to expose the photoresist, then transfers the pattern to the optical material by chemical etching. This technology is widely used to create microscopic arrays, waveguides, sensors and other optical components. With the continuous improvement of equipment, traditional photolithography techniques have achieved higher resolution and alignment accuracy \cite{kaganskiy2015advanced}. EUV lithography is an innovative method that uses an ultraviolet light source for exposure. EUV lithography has extremely high resolution and can handle more complex structures. It is widely used in the semiconductor industry but also has great potential in the production of optical devices, such as micro-lasers and optical communication components. 

Electron beam exposure (EBL) technology is a high-resolution preparation method that uses a precise electron beam to expose the photoresist. This method is often used to make optical devices that require extremely high precision and fine patterns, such as terahertz detectors and quantum dot arrays. EBL technology can achieve a precision of less than 100 nanometers, which is very beneficial for the fabrication of fine structures \cite{gschrey2015resolution}. 

In addition, nanoimprinting is an emerging method that reproduces micro- and nanostructures by applying pressure to the material surface. This technique is suitable for fabricating large-area structures such as nanoarrays and surface plasmonic resonators \cite{viheriala2010nanoimprint,Lan13}. This shows the potential of high-throughput manufacturing to increase production efficiency. 

In short, the diversity of lithography techniques offers a multitude of tools and methods for fabricating optical devices. Whether traditional lithography, EUV lithography, electron beam exposure or nanoimprinting, they play a key role in innovations in photonics and optoelectronics, driving optical equipment with high performance, high resolution and high reliability. As state-of-the-art lithography techniques and tooling continue to improve, it is natural that photonic devices will become more precise and complex, supporting further development of optical communications, neuromorphic computing and laser applications. 

\subsubsection{Etching}

Etching technology \cite{teo2005deep} is a key micro-nano processing method that is widely used for photonics and optoelectronics to accurately shape and regulate the surface and structure of optical materials. Reactive Ion Etching (RIE), for example, uses a combination of chemical and physical reactions to remove material from a substrate and is the simplest process that is capable of directional etching. A highly anisotropic etching process can be achieved in RIE through the application of energetic ion bombardment of the substrate during the plasma chemical etch. The RIE process thus provides the benefits of highly anisotropic etching due to the directionality of the ions bombarding the substrate surface as they get accelerated towards the negatively biased substrate, combined with high etch rates due to the chemical activity of the reactive species concurrently impinging on the substrate surfaces. 

The main application areas of etching include waveguides, gratings, lasers, optical modulators, etc. First, etching technology plays a crucial role in waveguide fabrication. Waveguide is a basic component of optical signal transmission. Complex waveguide structures can be produced through etching technology, including straight waveguides, curved waveguides and waveguide gratings, thereby achieving efficient guidance and coupling of optical signals \cite{fainman2013silicon}. The preparation of gratings and nanolasers is also inseparable from the etching technology. Gratings are used for spectroscopy and spectral analysis. By making micron-scale periodic structures (like holes or rods) on semiconductor materials, optical signals of specific wavelengths can be separated \cite{freude2004design}. Lastly, Lasers require high-precision preparation and delicacy, and etching can be used to precisely define the size and shape of the laser resonance cavity to ensure stable and efficient laser performance \cite{selvaraja2011loss}.

At the same time, optical modulators are indispensable devices in optical communications and optical signal processing, and are often prepared using etching technology. Etching techniques can be used to create gratings or modulation regions. By modulating the properties of these structures, modulation and control of optical signals, including amplitude, phase and frequency, can be achieved.
In short, the application of etching technology in optical devices provides key support for the development of photonics and optoelectronics. By precisely controlling optical structures, etching technology enables optical devices to achieve higher performance, greater scalability, and greater reliability, thereby driving advances in neuromorphic computing and photonic integrated circuits.

\subsubsection{3D printing technology }

3D printing technology \cite{campbell2011could}, also known as additive manufacturing, has emerged in the production of optical devices, bringing many potential applications and innovations to the fields of photonics and optoelectronics. This technology uses a layer-by-layer method of stacking materials, allowing the fabrication of complex optical structures while providing a high degree of customization and personalization. 

3D printing technology has many applications in the production of optical components. Complex optical components such as optical lenses, gratings, waveguides and reflectors can be precisely manufactured using 3D printing technology. This approach not only provides more flexible preparation but also reduces the number of optical components in the optical path, thereby reducing system complexity \cite{melzer20203d}, and 3D printing technology also plays a key role in manufacturing micro-optical devices. Micro-optical components such as micro-lenses and micro-lens arrays can be manufactured with high precision using 3D printing technology to meet the needs of micro-optical systems \cite{roques2022toward,fritzler20193d}. This is essential for applications such as micro cameras, biosensors and pico-projectors. In addition, optical waveguide fabrication also benefits from 3D printing, which can be used to create complex waveguide structures for efficient optical signal transmission. This is important for optical switches in data centers and optical communications because they deliver better performance and greater scalability. Therefore, by precisely controlling the arrangement and structure of materials, personalized photonic materials with special optical properties can be obtained \cite{jeong20203d,pyo20163d} for advanced computing tasks.

In short, 3D printing technology offers a new approach to manufacturing optical devices, with a high degree of flexibility and customization in the preparation process. This technology has made significant progress in the fabrication of optical components, micro-optical devices, optical waveguides and photonic materials, providing more opportunities for the development of photonics and optoelectronics. As 3D printing technology continues to develop and innovate, we will see its broader adoption in neuromorphic engineering. 

\subsubsection{Challenges and Future Directions}

While remarkable progress has been made in photonic fabrication, several challenges remain:

\textit{Integration}: efficient integration of various photonic components into functional systems, such as a PIC, is a complex task that requires interdisciplinary collaboration. Some existing integration methods include: heterogeneous, monolithic, hybrid integration etc.

\textit{Scalability}: many manufacturing methods need to be adapted to large-scale production to meet the growing demand for photonic devices in the post Moore era. Current photonic manufacturing has relatively poor scalability compared to electronic ICs.

\textit{Materials innovation}: the development of new materials with tailored optical properties will expand the design possibilities of photonic devices. Core semiconductor materials, besides the typicel Silicon, are Gallium Arsenide, Lithium Niobate, Indium Phosphide, Silicon Nitride, and Transition-metal dichalcogenide (TMDCs).

\textit{Cost-Efficiency}: reducing the cost of fabrication techniques, especially for lithography, 3D printing and self-assembly, is essential for widespread adoption of photonic devices. This could also call for cheaper fabrication tools and equipment.

\textit{Hybrid Platforms}: combining different fabrication methods can offer unique advantages, leading to hybrid photonic platforms. Common semiconductor platforms include Complementary metal–oxide–semiconductor (CMOS), carbon systems including carbon nanotubes and graphene, and the latest magnetoelectric spin–orbit (MESO) \cite{manipatruni2019scalable} etc.

In conclusion, photonics has witnessed substantial advancements in fabrication platforms, enabling the creation of intricate and efficient photonic devices. Lithography, etching, self-assembly, and 3D printing have all played pivotal roles in shaping the future of photonics. With continued research and innovation, photonic devices are poised to revolutionize a wide range of applications, from telecommunications and neuromorphics to medical imaging and AI, ushering in a new era of photonics on the nanoscale.

\begin{table}[htbp]
\footnotesize
\caption{A summary of recent milestone photonic neuromorphic paradigms for AI and deep learning applications. Sources are listed chronologically. Only architectures that are CMOS-compatible chips are given compute density. Cost refers to the estimated fabrication cost per unit chip area according to the fabrication platform and devices used. \cite{feldmann2021parallel} and \cite{xu202111}, adopting similar technologies, were published on the same day. For pictorial illustration of each work, refer to Figure 3. NA=not available or not calculated. SOI=silicon-on-insulator.}
\centering
\rowcolors{1}{yellow}{yellow}
\begin{tabular}{| p{0.08\linewidth} | p{0.15\linewidth} | p{0.1\linewidth} | p{0.11\linewidth} | p{0.06\linewidth} | p{0.1\linewidth} | p{0.1\linewidth} | p{0.1\linewidth}|}
\hline
\textbf{Name abbrv.}  & \textbf{Technologies \& methods} &\textbf{Energy efficiency \tiny ($TOP/J$)}  & \textbf{Compute density \tiny ($TOP/mm^{2}/s$)} &\textbf{Est. cost \tiny $/mm^{2}$} & \textbf{Input encoding} &\small \textbf{Implementation platform} & \textbf{Reference} \\ \hline

PNP & Mach–Zehnder interferometers, silicon photonics, photodiode, phase shifter & NA & NA & $\$\$$ & Laser optical pulses & CMOS-compatible photonic chip & \cite{Shen_Harris_Skirlo_Prabhu_Baehr-Jones_Hochberg_Sun_Zhao_Larochelle_Englund_et} 2017 Nature Photonics \\ \hline

$D^2NN$ & 3D printed lenses and optical diffraction & NA&  NA  & $\$\$$ & Optical image signal & Free space \& Bench-top & \cite{Lin_Rivenson_Yardimci_Veli_Luo_Jarrahi_Ozcan_2018} 2018 Science\\ \hline

AONN & Spatial light modulator, Fourier lens, laser-cooled atom & NA& NA & $\$\$$ & Optical image signal & Free space \& Bench-top &\cite{Zuo_Li_Zhao_Jiang_Chen_Chen_Jo_Liu_Du_2019} 2019 Optica\\\hline

Spiking neurosynaptic network & Phase change material, micro-resonator, and wavelength division multiplexing & NA & NA  & $\$$ & Laser optical pulses & CMOS-compatible photonic chip &\cite{feldmann2019all} 2019 Nature \\ \hline

Photonic tensor core & Phase change material, soliton microcombs, SiN micro-resonator, and wavelength division multiplexing& 0.4 & 1.2&  $\$$ & Soliton frequency comb& CMOS-compatible photonic chip & \cite{feldmann2021parallel} Jan 2021 Nature \\ \hline

Optical convolutional accelerator & Soliton microcombs, micro-resonator, Mach–Zehnder modulator, EDFA, and time-wavelength interleaving & 1.27 & 8.061 & $\$\$$ & Electrical waveform & CMOS-compatible photonic chip & \cite{xu202111} Jan 2021 Nature \\ \hline

PDNN & PIN attenuator, SiGe photodiodes, grating coupler, and microring modulator & 2.9& 3.5 & $\$\$$ & Optical image signal & CMOS-compatible SOI Photonic chip &\cite{ashtiani2022chip} 2022 Nature \\ \hline

PAIM & Meta-surface, optical diffraction, and FPGA & NA & NA & $\$\$\$$ & Optical image signal & Free space \& Bench-top& \cite{liu2022programmable} 2022 Nature Electronics\\ \hline

VCSEL-ONN & VCSEL, diffractive optical element, and optical fanout & 142.9 & 6 & $\$\$$ & Amplitude or phase of VCSEL & CMOS-compatible photonic chip &\cite{chen2023deep} 2023 Nature Photonics \\ \hline


\end{tabular}
\end{table}

\section{Recent Advances in Photonic Neuromorphic Computing}

Figure 9 systematically compares biological (human brain), electronic, and photonic neuromorphic computing systems across the board. Key metrics / FoM being compared include: energy consumption, parallelism, latency, noise-level, computing speed, 3D or 2D topology, and learning capacity. From Figure 9 we see that photonic systems outperform electronic counterparts in all the aspects above and even match the energy consumption level of human brains. Similar trends can be observed in Figure 5.

\begin{figure}[hbtp]
\centering
\includegraphics[width=.85\textwidth]{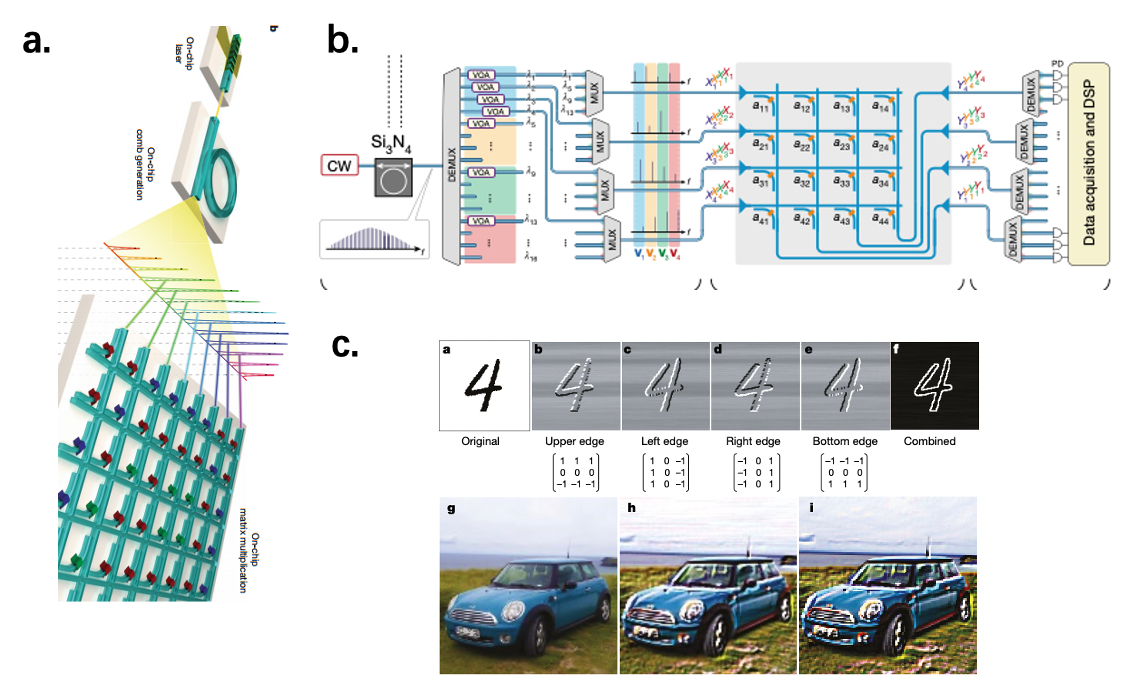}
\caption{Convolutional neural networks (CNN) results demonstrated in the photonic tensor core paper \cite{feldmann2021parallel}. a. Schematic of the photonic tensor core AI accelerator. b. Circuit diagram of the multiplexed all-optical convolutional tensor core. The input vectors are generated from a photonic frequency microcomb driven by a continuous-wave (CW) laser using wavelength division multiplexers. Right side involves the MAC unit (waveguides and PCMs). c. Convolution using sequential MVM operations on images of a handwritten digit 4 and a car. Using a high degree of parallelism, the processing time in c is reduced by a factor of 4. Reproduced with permission.\cite{feldmann2021parallel} Copyright 2021, Springer Nature.
}\label{fig:1}
\index{figures}
\end{figure}

Table 1 more systematically tabulates and compares the recent milestone photonic neuromorphic paradigms for AI applications proposed by well-established institutions and groups around the globe. This table chronologically lists the technologies and methods used for achieving these representative photonic architectures, including the materials, devices, implementation platforms, encoding methods, and associated fabrication costs. Simultaneously, the performance of each architecture is reported as extracted from their individual papers, including the FoMs such as energy efficiency ($TOP/J$) and compute density ($TOP/mm^2/s$), where the former stands for tera ($10^{12}$) operations per joule while the latter tera operations per squared millimeter per second. The goal of state-of-the-art neuromorphic systems is achieving peta-level ($10^{15}$) in both energy efficiency and compute density. From Table 1 it's clear that while both energy efficiency and compute density have steadily risen with time, existing efforts struggle to meet the peta ($10^{15}$) threshold and they are still stuck on the tera-level in both FoMs for the most part. A comprehensive plot of these FoMs among various electronic and photonic neuromorphic systems is illustrated in Figure 5, which shows that photonic systems already outperform electronic counterparts in these FoMs. From the perspective of engineering, in Table 1 we also tabulate the estimated cost of fabrication for each work, aiming to compare and quantify how photonic neuromorphic systems' fabrication cost evolves with time or varies among different structures. The fabrication platform and devices used play a major role in the cost. Next, we observe that in the systems of \cite{ashtiani2022chip} and \cite{chen2023all}, the input encoding (optical image signals) directly comes from sensing targets, the energy cost and encoding rates of which are not taken into account, whereas in \cite{feldmann2021parallel}, \cite{xu202111}, and \cite{chen2023deep}, input encoding (e.g., frequency comb and electrical waveform) is a major difficulty to improve the overall system performance. Finally, we distinguish architectures that are integrated CMOS-compatible photonic chips from those that are free space bench-top implementations. Only those that are integrated chips are given the compute density  (because it doesn't make sense to calculate those per-unit-area FoMs for free-space implementations) and we believe chip implementations offer much better computing and energy performances compared to free-space ones in the long run. As such, this table aims to summarize and contextualize existing approaches to photonic neuromorphic computing and provide the readers a bigger picture of the state-of-the-art development of this field up to now. Below, we introduce several of these representative works in detail.

\begin{figure}[hbtp]
\centering
\includegraphics[width=.85\textwidth]{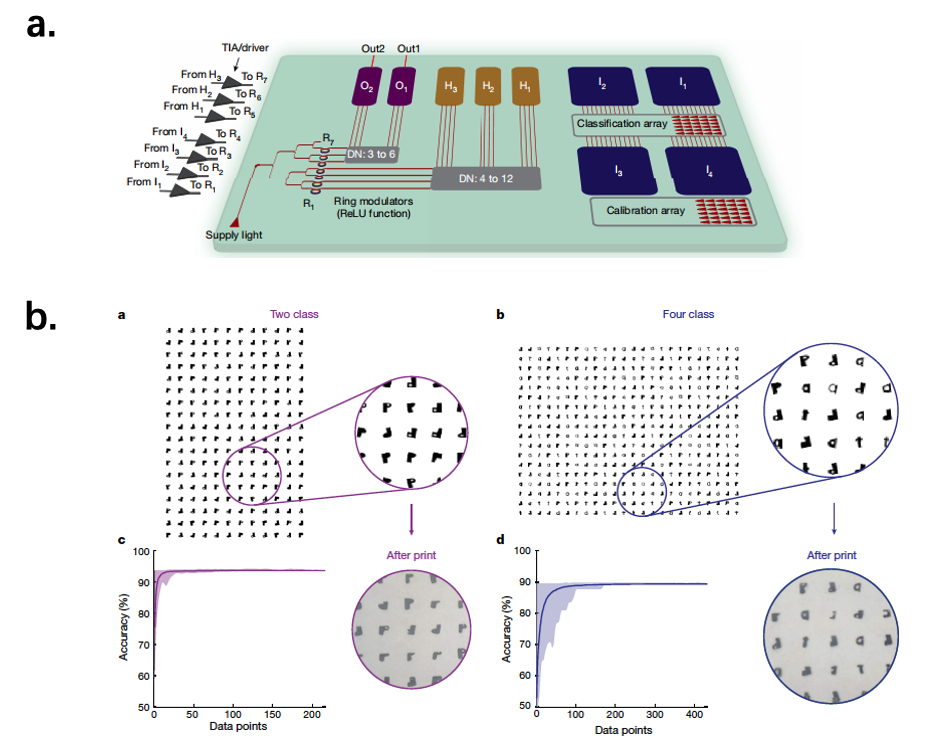}
\caption{The implemented photonic classifier chip PDNN \cite{ashtiani2022chip}. a. The top-level block diagram of PDNN and the structure of
an implemented N-input photonic neuron, in which the weights of N optical
input signals are adjusted using optical PIN attenuators and summed after
photodetection using parallel PDs. An optical MRM realizes the ReLU non-linear activation function. b. Image classification demonstration, with the two-class dataset consisting of letters ‘p’ and ‘d’ (left) and four-class dataset consisting of ‘p’, ‘d’, ‘a’ and ‘t’ (right). Classification accuracy of both classes are plotted. Reproduced with permission.\cite{ashtiani2022chip} Copyright 2022, Springer Nature.
}\label{fig:1}
\index{figures}
\end{figure}

In Figure 13, parallel, fast, and efficient convolution operations are realized by a photonic tensor core \cite{feldmann2021parallel}, which is enabled by technologies such as phase change material (PCM), soliton microcombs, SiN micro-resonator, and wavelength division multiplexing (WDM) etc. The tensor core can be likened to the optical equivalent of an ASIC, achieving highly parallelized photonic in-memory computing through PCM memory arrays and optical frequency combs based on photonic chips. The computation involves measuring the optical transmission of reconfigurable and non-resonant passive components, capable of operating at a bandwidth exceeding 14 GHz, constrained only by the speeds of modulators and photodetectors. In this study, the tensor core demonstrated a performance level of 2 tera-MAC operations per second (tera being a trillion). Notably, the convolutional operation, being a passive transmission measurement, theoretically allows calculations at the speed of light with very low power consumption (approximately 17 fj per MAC), limited in experiments only by modulation and detection bandwidths. Recent advancements in hybrid integration, including soliton microcombs at microwave line rates, ultralow-loss \ce{Si3N4} waveguides, and high-speed on-chip detectors and modulators, pave the way for the potential full CMOS wafer-scale integration of the photonic tensor core with silicon photonics in this innovative paradigm.

In Figure 14, the PDNN paper \cite{ashtiani2022chip} demonstrated the first end-to-end PDNN photonic classifier chip that performs sub-nanosecond image classification through computation by propagation of optical waves, eliminating
the need for an image sensor, digitization and large memory modules. PDNN processes optical waves that reach the on-chip pixel array as they traverse layers of neurons. Within each neuron, optical linear computation takes place, and the non-linear activation function is achieved opto-electronically, resulting in a classification time of less than 570 ps—comparable to a single clock cycle of cutting-edge digital platforms. Initially, a set of 500-μm-long PIN current-controlled attenuators is employed to individually tune the optical power in each input nanophotonic waveguide of the neuron. By forward biasing the PIN junction and injecting carriers, the power of the optical wave (i.e., the signal weight) for each neuron input can be adjusted. The outputs of attenuators are photodetected using SiGe photodiodes (PDs), and the resultant photocurrents are combined to generate the weighted sum of neuron inputs. A uniformly distributed supply light ensures a consistent per-neuron optical output range, facilitating scalability to large-scale PDNNs. Successful demonstrations include two-class and four-class classification of handwritten letters with accuracies exceeding 93.8\% and 89.8\%, respectively. Notably, the photonic classifier chip offers low energy consumption and ultra-low computation time, presenting revolutionary potential in applications like event-driven and salient object detection in autonomous driving. It can function as a stand-alone classifier or in conjunction with electronic processors, benefiting from the sub-nanosecond classification capability of the PDNN chip. Lastly, the PDNN chip is implemented in the silicon-on-insulator (SOI) process. SOI fabrication processes offer monolithic integration of electronic and photonic devices.

\begin{figure}[hbtp]
\centering
\includegraphics[width=.85\textwidth]{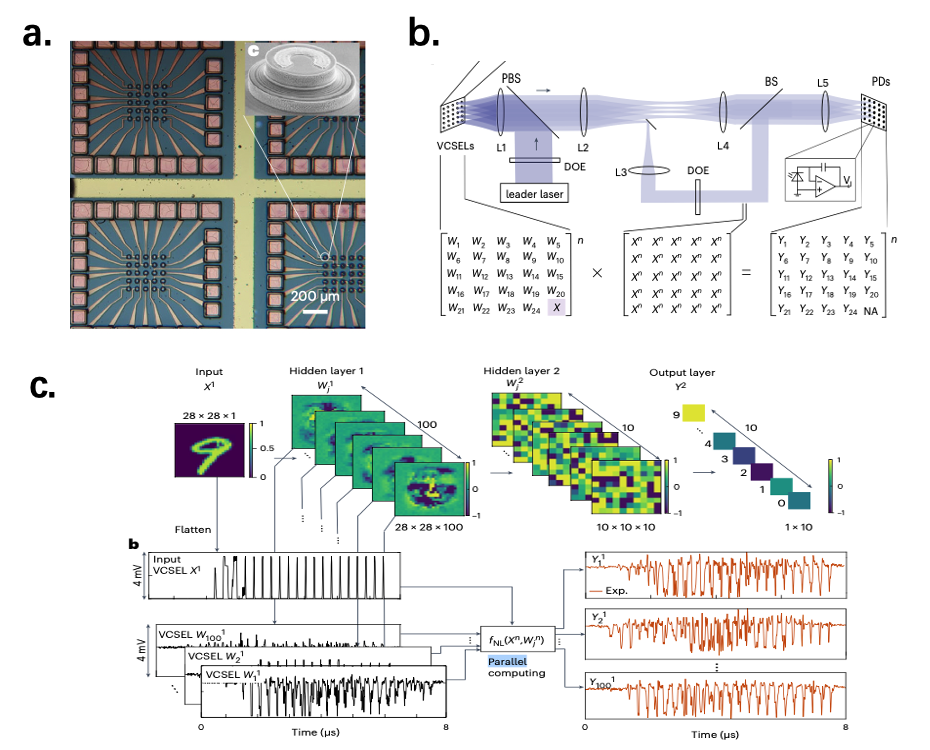}
\caption{Parallel matrix-vector multiplication realized by VCSEL-ONN \cite{chen2023deep}. a. Fabricated VCSEL arrays. Arrays of 5 × 5 wire-bonded VCSELs on a GaAs substrate and an SEM image of a VCSEL emitter. b. Proposed architecture with 3D connectivity and photonic integration. In a 2D VCSEL array, the center unit is used as the axon and the others as weight VCSELs. The axon beam is fanned out to j copies, each overlapping a weighted beam onto a photodetector, generating photon currents corresponding to the homodyne product of the two laser fields. The input VCSEL is separated from the beam arrays using a beam magnifier (L1 and L2) and D-shaped mirror. c. Benchmarking of machine learning inference with VCSEL-ONN using MNIST. The input image in layer 1 is flattened and encoded in time steps to the phase of the Xn VCSEL. The weight matrix with 100 vectors is encoded to 100 individual weighting VCSELs. Parallel multiplication results in MVM from 100 readout channels. Reproduced with permission.\cite{chen2023deep} Copyright 2023, Springer Nature.
}\label{fig:1}
\index{figures}
\end{figure}

In Figure 15, the VCSEL-ONN paper \cite{chen2023deep} experimentally demonstrates a spatial-temporal-multiplexed ONN system that mitigates several key challenges faced by existing ONNs: low electro-optic conversion rate, large device footprint, and long latency. This paper explores neuron encoding using micrometer-scale vertical-cavity surface-emitting laser (VCSEL) arrays, characterized by efficient electro-optic conversion (< 5 attojoules per symbol) and a compact footprint (< 0.01 $mm^2$ per VCSEL array, not including the full setup with additional free-space components). Employing homodyne photoelectric multiplication enables matrix operations at the quantum-noise limit, accompanied by detection-based optical nonlinearity featuring an instantaneous response. The VCSEL-ONN architecture comprises N layers, where each layer performs a MVM followed by a nonlinear activation function, mimicking the biological neurons' "axon-synapse-dendrite" architecture. VCSEL-ONN encodes the input vector in i time steps to the amplitude or phase of a coherent laser oscillator (referred to as 'axon'), with its beam dendritically fanning out to j copies for parallel processing. With three-dimensional neural connectivity, this system achieves an energy efficiency of 7 fj per operation and a compute density of 6 $teraOP mm^{-2}s^{-1}$, representing 100-fold and 20-fold improvements, respectively. Benchmarking on the MNIST dataset for digit classification yields an accuracy of 93.1\% (over 98\% of ground truth). The system is anticipated to perform MAC operations with an efficiency of ~50 aJ/OP, primarily limited by memory access rather than optical energy consumption. It should also be noted that the full VCSEL-ONN includes some components in a free-space setup, which means that we are still some distance away from a fully integrated photonic neuromorphic system.

Although integrated photonics has been proven successful in photonic neural networks for neuromorphic computing, it has relatively poor scaling and the size of matrices remains small at the moment. Therefore, besides the above milestone integrated photonic chips, we introduce several recent free-space ONNs that enable nonlinearity operations. 

Free-space optical implementations, in particular, hold promise for significant speed enhancements and reduced energy consumption in the deep learning realm, thanks to its inherent parallelism. However, realizing the essential nonlinear component of DNN in free-space optics presents challenges, limiting the platform's potential. Moreover, achieving parallel nonlinear activation for each data point adds complexity to preserving the advantages of linear free-space optics. To that extent, people have introduced a free-space optical neural network featuring diffraction-based linear weight summation and nonlinear activation facilitated by the saturable absorption of thermal atoms \cite{ryou2021free}. Specifically, they exploit the saturable absorption behavior of room-temperature rubidium atoms housed in a vapor cell and observed the nonlinearity in a single pass without any cavity, which allows point-by-point nonlinear activation of an incident image. In the paper, they demonstrated image classification of handwritten digits using only a single layer. Compared to a linear model, their optical nonlinearity contributes to a 6\% improvement in classification accuracy \cite{ryou2021free}. This platform maintains the significant parallelism inherent in free-space optics, even with physical nonlinearities, paving the way for the broader adoption of ONNs in neuromorphic computing. Another work \cite{Zuo_Li_Zhao_Jiang_Chen_Chen_Jo_Liu_Du_2019} utilized a similar approach to this one.

In the realm of image sensing, the process of obtaining an object's location or shape involves analyzing captured images in digital computers. A novel approach to image sensing involves using optical systems not for traditional imaging but as encoders that compress images optically into low-dimensional spaces by extracting key features. However, the effectiveness of these encoders is typically constrained by the linearity.
In a recent study \cite{wang2023image}, researchers introduced a free-space, nonlinear, and multilayer ONN encoder for image sensing. This ONN encoder utilizes an image intensifier as an optical-to-optical nonlinear activation (OONA) function. The ONN encoder comprises an optical matrix–vector multiplier unit, an OONA (nonlinear) unit, a second optical matrix– vector multiplier, and a camera. Compared to similarly sized linear optical encoders, this nonlinear ONN demonstrates superior performance across various tasks, including computer vision, flow-cytometry image classification, and identifying objects in a 3D-printed real scene. Particularly for computer vision tasks employing incoherent broadband illumination, their paradigm allows for significant reductions in camera resolution requirements and electronic post-processing complexity. Overall, employing free-space ONNs holds promise for image-sensing applications that can operate accurately with fewer pixels, fewer photons, lower power consumption, increased throughput, and reduced latency.

\section{Emerging Areas in Photonics for Neuromorphic Computing}

\subsection{Emerging light sources and their potential impact}
Emerging nanophotonic technologies are leading the way in scientific and engineering progress, offering innovative approaches to observe the world through precise control of light at the nanoscale. In the wake of the seminal discovery of the quantum Hall effect, the physics community has embarked on a journey marked by significant advancements in the realm of topological insulators (TI). For instance, in the presence of a magnetic field within the microwave frequency range, researchers observed the unidirectional transmission of electromagnetic fields, revealing their remarkable anti-scattering characteristics and robustness \cite{Wang_Chong_Joannopoulos_Soljačić_2009}. This unique attribute renders topological insulators exceptionally suitable for its incorporation in next-gen nanophotonic devices. Simply put, TI enhances the fault tolerance and robustness of nano-scale on-chip integrated optical systems, while enabling functionalities that were previously challenging to attain with conventional optical components.

\begin{figure}[hbt!]
\centering
\includegraphics[width=.9\textwidth]{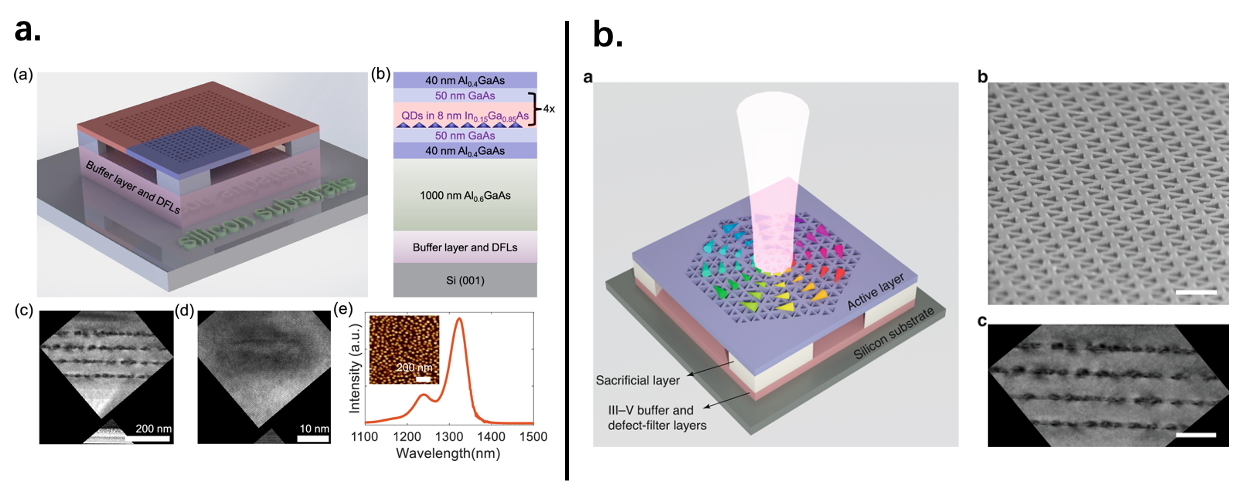}
\caption{State-of-the-art topological photonic lasers. a. Schematic structure of the fabricated topological corner state nanolasers monolithically integrated on a CMOS-compatible silicon substrate \cite{Zhou_Ma_Tang_Li_Martin_Baron_Liu_Chen_Sun_Zhang_2022}. High-resolution TEM images of the four stacked InAs/GaAs QDs and a single QD in the active region, respectively, are shown. Also shows the measured PL spectra of as-grown InAs/GaAs QDs on silicon. b. Conceptual illustration of a topological Dirac-vortex microcavity laser fabricated on a silicon substrate \cite{Ma_Zhou_Tang_Li_Zhang_Xi_Martin_Baron_Liu_Zhang_et}. The photonic crystal structure was defined in the active layer and suspended by partially removing the sacrificial layer. The III–V buffer and defect-filter layers were carefully optimized to minimize the effects of lattice mismatch between the III–V materials and silicon substrate. Tilted-view SEM image of the fabricated topological laser and cross-sectional bright-field TEM image of the active layer containing four-stack InAs/InGaAs QD layers are shown. Panels reproduced with permission from: a. Reproduced with permission.\cite{Zhou_Ma_Tang_Li_Martin_Baron_Liu_Chen_Sun_Zhang_2022} Copyright 2022, American Chemical Society. b. Reproduced under the terms of the Creative Commons Attribution 4.0 International License (http://creativecommons.org/licenses/by/4.0/).\cite{Ma_Zhou_Tang_Li_Zhang_Xi_Martin_Baron_Liu_Zhang_et} Copyright 2023, the Authors and Springer Nature.}\label{fig:1}
\index{figures}
\end{figure}

Recently, the emergence of topological photonics marks a profound departure from conventional optical paradigms, introducing the concept of "topological invariants" and "topologically protected state" to the domain of optics. This innovative approach delves into the intriguing premise that wave functions of electromagnetic fields remain impervious to alteration amidst fluctuating geometric structures. Consequently, topological photonic devices began to widely emerge, distinguished by their exceptional optical performance attributes and immunity to perturbations. These include an inherent robustness, a remarkable capacity to resist interference, the effective suppression of photon backscattering, and the realization of a notably high free spectral range (FSR). All of these above attributes render topological photonics an attractive candidate for building integrated on-chip light sources for neuromorphic computing, such as those lasers employed in \cite{Shen_Harris_Skirlo_Prabhu_Baehr-Jones_Hochberg_Sun_Zhao_Larochelle_Englund_et}, \cite{feldmann2019all}, \cite{feldmann2021parallel}, and especially \cite{chen2023deep}.


In non-Hermitian systems, nanoscale topological photonic crystals, due to their remarkable robustness and anti-scattering properties, show great potential in laser technology. By harnessing the unique characteristics of boundary states, bulk states, corner states (Fgure 16a), and other features inherent to topological photonic crystals, light sources with diverse optical properties can be realized \cite{Bahari_Ndao_Vallini_El_Amili_Fainman_Kanté_2017,Shao_Chen_Wang_Mao_Yang_Wang_Wang_Hu_Ma_2020,Zhang_Xie_Hao_Dang_Xiao_Shi_Ni_Niu_Wang_Jin_et,Zhou_Ma_Tang_Li_Martin_Baron_Liu_Chen_Sun_Zhang_2022}. The photonic band structure at the Dirac point brings distinct property of zero refractive index, and combined with the transmission property of topological photonic crystals, a kind of vortex state laser can be designed by precisely controlling the phase of the electromagnetic field \cite{Ma_Zhou_Tang_Li_Zhang_Xi_Martin_Baron_Liu_Zhang_et,Yang_Li_Gao_Lu_2022} (Figure 16b). This breed of laser, allowing for meticulous control of the light field distribution at the nanoscale, boasts an impressively high free spectral range and surface emission characteristics, thus emerging as a highly competitive light source for neuromorphic computing. This novel Dirac-vortex laser \cite{Ma_Zhou_Tang_Li_Zhang_Xi_Martin_Baron_Liu_Zhang_et} has the potential to outperform the VCSEL lasers utilized in \cite{chen2023deep} in terms of computing speed, energy efficiency, electro-optic conversion rate, and stability. In addition, to facilitate the transition of topological photonics into practical applications, researchers conceived topological quantum cascade lasers powered by terahertz waves, thereby significantly amplifying the potential of topological lasers for neuromorphic computing \cite{Han_Chua_Zeng_Zhu_Wang_Qiang_Jin_Wang_Li_Davies_et,Zeng_Chattopadhyay_Zhu_Qiang_Li_Jin_Li_Davies_Linfield_Zhang_et}.

Within Hermitian topological systems, valley photonic crystals were used to achieve lossless topological photonic insulator waveguides, even when subjected to sharp bends \cite{Shalaev_Walasik_Tsukernik_Xu_Litchinitser_2019}. Furthermore, researchers delved into the exceptional properties of topological slow-light waveguides \cite{Arregui_Gomis-Bresco_Sotomayor-Torres_Garcia_2021} and polarization beam splitters \cite{He_Zhang_Zhang_Wang_Zhang_2021}, all of which can be integrated in neuromorphic computing chips in conjunction with the aforementioned topological lasers. Importantly, the topological protection after laser emission will still be held as long as the coupling components (such as waveguides) belong to the same class of topological photonic structure as the laser. Topological photonic components also can be integrated into a CMOS-compatible silicon-on-insulator (SOI) platform \cite{He_Liang_Yuan_Qiu_Chen_Zhao_Dong_2019}, exemplifying the immense potential of integrated topological photonics within the field of neuromorphic computing. 


\begin{figure}[hbt!]
\centering
\includegraphics[width=.7\textwidth]{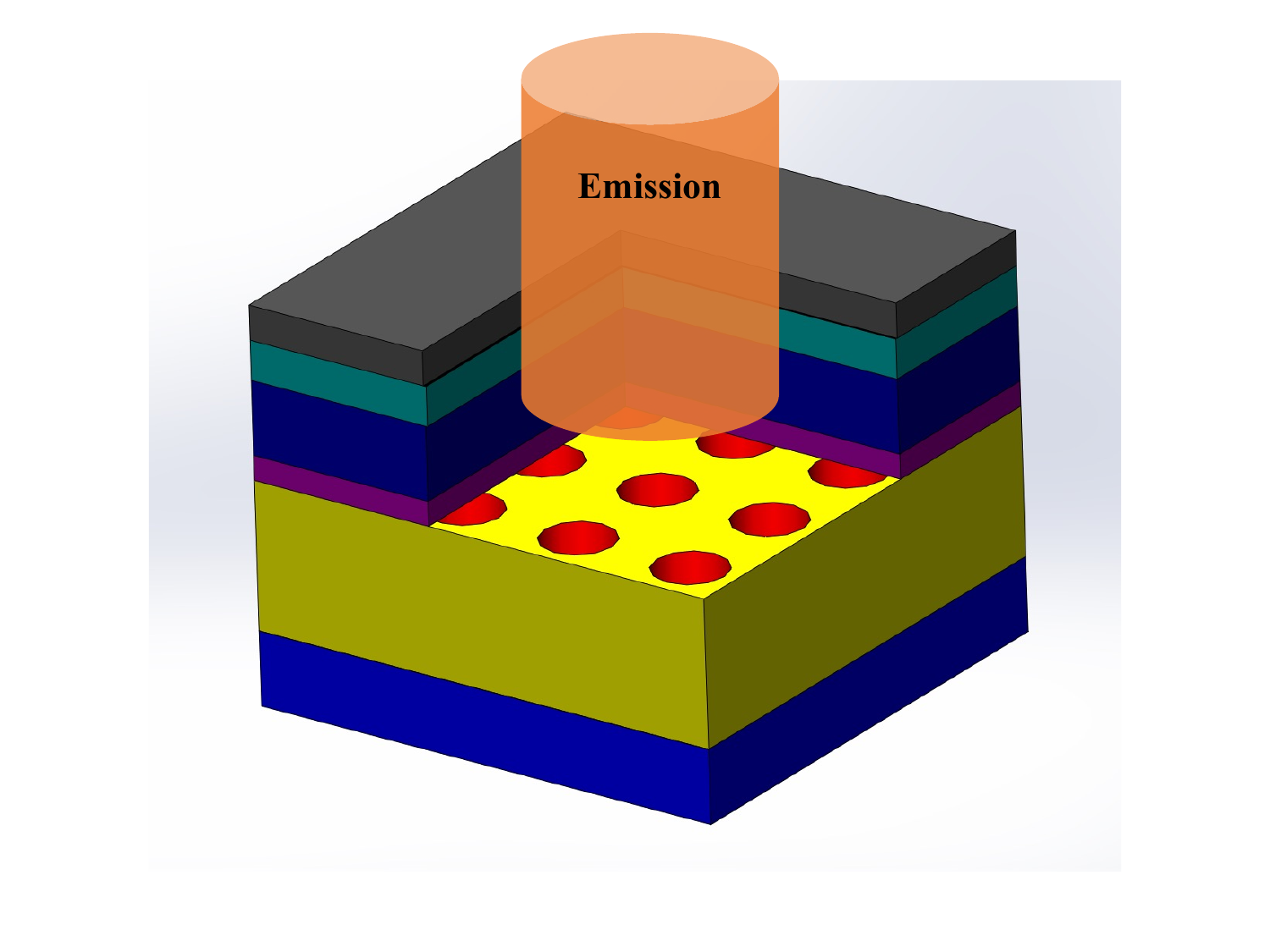}
\caption{Schematic of a standard PCSEL device with circular air holes in the PhC layer. Multiple epitaxial layers and substrates are symbolically shown, with their names omitted for simplicity. Laser beam is emitted vertically.}\label{fig:1}
\index{figures}
\end{figure}

In addition to topological lasers, Photonic Crystal Surface Emitting Lasers (PCSELs) \cite{hirose2014watt,yoshida2019double,li:hal-04175312,noda2017photonic} (Figure 17) are a novel type of laser that combines the advantages of photonic crystals (PhCs) \cite{quan2010photonic} and Vertical Cavity Surface Emitting Lasers (VCSELs) \cite{chang2000tunable}. PhCs are artificially engineered structures with periodic changes in refractive index in one, two, or three dimensions, creating a bandgap that restricts the propagation of light in specific frequency ranges. VCSELs are lasers that emit light perpendicular to the surface of a semiconductor, facilitating efficient coupling with optical fibers and other optical components. PCSELs merge these technologies to create lasers with multiple benefits compared to traditional lasers, offering the best of both worlds.

The fundamental structure of a PCSEL includes a PhC layer, an active layer, additional cladding layers, substrates, p-n junctions, and electrodes at the ends. The PhC layer mainly functions as a resonant cavity, while the active layer, typically composed of III-V materials (such as InP/InGaP, GaAs/InGaAs/AlGaAs, GaN/InGaN, etc.), is positioned in the middle of the PCSEL to induce laser emission when a population inversion of charge carriers is achieved upon reaching a certain threshold \cite{sze2021physics,hirose2014watt}. Population inversion occurs when there are more electrons in higher energy states than in lower energy states, allowing for stimulated emission of photons when an electrical current is applied. When an electrical pumping current is introduced into the active layer material, it emits laser light that is effectively confined and amplified within the resonance cavity. Furthermore, the active layer can contain quantum dots or quantum wells, which increase the recombination rate of spontaneous emission, significantly enhancing the lasing effect. Consequently, PCSELs have superior characteristics than conventional VCSELs.

\subsection{Emerging silicon-on-insulator (SOI) paradigm and its potential impact}

Monolithically integrated microcavity lasers on heterogeneous epitaxy-enabled SOI substrates (Figure 17) is a promising method to achieve in-plane light source coupling within photonic integrated circuits (PIC). The SOI paradigm can greatly improve the integration density and scalability while reducing the cost of PICs,   while at the same time offering monolithic integration of electronic and
photonic devices on the same chip. Besides, it can help overcome the common large lattice mismatch, large diffusivity of indium adatoms, and the thick buffer layer issues of III-V materials monolithically grown on Si substrate.  In short, SOI is expected to play a significant role in light source integration of PICs for short-distance optical communication and data centers. Faster speed, ultrasmall footprint, low power consumption, low cost, and CMOS compatible fabrication ensure the significant position of SOI-based PICs in data communication network and optical computing.

\begin{figure}[hbt!]
\centering
\includegraphics[width=.95\textwidth]{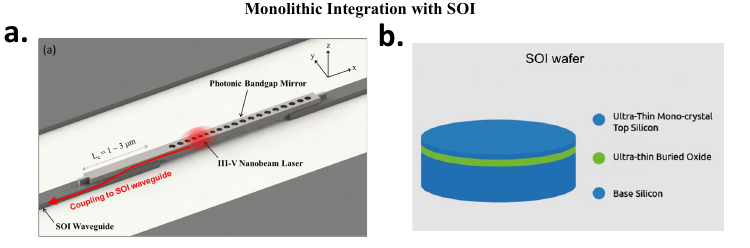}
\caption{a. Schematic diagram of a PhC nanobeam laser grown on silicon-on-insulator (SOI) substrate, monolithically integrated with Si waveguides, electrodes, and other on-chip devices. Reproduced with permission.\cite{lee2017printed} Copyright 2017, American Chemical Society. b. side view of a typical SOI wafter, where the buried oxide is Silica. }\label{fig:1}
\index{figures}
\end{figure}

\subsection{Emerging optical encoder-ANNs and their potential impact}
The recent advancements in utilizing light for conducting large-scale linear operations in parallel have sparked numerous demonstrations of optics-integrated artificial neural networks (ANNs). Nevertheless, a definitive system-level superiority of optics over entirely digital ANNs has yet to be firmly established. While optical systems excel in efficiently executing linear operations, the absence of nonlinearity and signal regeneration necessitates high-power and low-latency signal transmission between optical and electronic systems. Moreover, substantial power requirements for lasers and photodetectors, often overlooked in energy consumption calculations, further complicate the picture.

In this context, instead of merely translating conventional digital operations into optics, people have developed a hybrid optical-digital ANN through utilizing a meta-optical encoder \cite{huang2024photonic}. This hybrid model operates using incoherent light, making it compatible with ambient light conditions. By maintaining consistent latency and power levels between a fully digital ANN and our hybrid optical-digital ANN, they have identified a regime characterized by low power and latency. Within this regime, an optical encoder outperforms a purely digital ANN in terms of classification accuracy on MNIST dataset. Their estimates suggest that the optical encoder facilitates operation rates of over 10 kHz for a hybrid ANN, consuming only 23 mW of power. However, within this regime, the overall classification accuracy is slightly lower compared to what can be achieved with higher power and latency settings.

These findings suggest that optics can offer advantages over digital ANNs in scenarios where prioritizing lower power consumption and latency is paramount, even if it entails some compromise on overall performance \cite{huang2024photonic}.


\section{Remaining Challenges and Future Directions}

\subsection{Strategies for improving device performance, scalability, and reliability}

\begin{table}[htbp]
\small
\caption{Strategies for improving photonic device performance, scalability, and reliability. The realm of photonics plays a vital role in improving data transfer speeds in optical communication, thus achieving scalability in photonic components is a pressing goal. Device reliability is paramount, particularly in critical applications like aerospace and healthcare. Strategies to enhance reliability encompass both design and materials.}
\bigskip
\centering
\rowcolors{1}{yellow}{yellow}
\begin{tabular}{|p{0.31\linewidth} | p{0.31\linewidth} | p{0.31\linewidth}|}
\toprule
\textbf{Performance}  & \textbf{Scalability} & \textbf{Reliability}\\ \hline\hline 
Advanced Transistor Architectures: The introduction of novel transistor architectures, such as FinFETs and gate-all-around (GAA) nanosheets, has significantly improved the performance of microprocessors. These designs offer enhanced control of the current flow, reduced leakage, and higher switching speeds \cite{srivastava202311}. & Photonic Integrated Circuits (PICs): PICs have emerged as a key strategy to integrate multiple optical modules onto a single chip, reducing size and increasing functionality \cite{errando2019mems}. Advances in PIC design and fabrication have improved the performance of devices like optical transceivers and lasers. & Redundancy and Error Correction: in electronic systems, redundancy and error correction techniques are employed to improve reliability. These methods ensure that even in the presence of defects or failures, the system continues to function correctly \cite{fernandez2022error}. \\ \hline

High-Mobility Materials: The integration of high-mobility materials, like gallium nitride (GaN) and indium gallium arsenide (InGaAs), has led to faster and more efficient optoelectronic devices. GaN power devices, for example, have found applications in power electronics and RF amplifiers 
\cite{zhou2013fabrication}. & Meta-surfaces: Meta-surfaces consist of subwavelength nanostructures that manipulate light in novel ways. These structures are integral in creating flat optical components, enabling highly compact and scalable photonic devices \cite{hu2021review,zhao2021recent}. & Photonic Error Correction:  There have been experimental demonstrations of deterministic real time training and error correction in integrated photonic systems \cite{filipovich2022silicon,wu2023lithography,pai2023experimentally} to realize efficient deep learning in photonic neural networks (via in situ training).  \\ \hline

Quantum Devices: Quantum devices, such as quantum-dot transistors and quantum cascade lasers, offer unique properties that can revolutionize electronics and photonics. Quantum computing, for example, has the potential to solve complex problems at unprecedented speeds \cite{pelucchi2022potential,laucht2021roadmap}. & Additive Manufacturing: Additive manufacturing, including 3D printing, allows for the rapid and cost-effective production of complex components. It has applications in aerospace, healthcare, and customized electronic packaging \cite{khanpara2020additive,mehrpouya2019potential}; Monolithic 3D Integration: Monolithic 3D integration is a disruptive manufacturing technique that enables the stacking of active devices in the vertical direction (also known as epitaxy), enhancing performance and reducing interconnect lengths 
\cite{dhananjay2021monolithic,zhang2022recent,chen2023challenges,chaourani2019sequential}. & Reliability Testing: Rigorous reliability testing, including accelerated life tests, is conducted to identify and mitigate potential failure modes. This process ensures that devices meet their specified lifetime and performance criteria. \cite{girard2020survey}. \\ \hline
\end{tabular}
\end{table}

As neuromorphic technology continues to advance and demands are getting higher, the performance, scalability, and reliability of photonic devices become increasingly critical. Achieving higher efficiency, speed, and functionality is essential to meet the ever-growing demands of modern computing tasks. This section (Table 2) will address recent developments and strategies implemented to enhance device performance, scalability, and reliability in neuromorphic photonics, where engineers and scientists are exploring a multitude of strategies. These encompass device engineering techniques, the incorporation of novel materials, and advancements in manufacturing processes. These strategies are integral in fields as diverse as semiconductor optoelectronics, photonics, and integrated systems. For instance, \cite{wu2023lithography} proposed a reconfigurable integrated photonic processor that performs in situ training of vowel recognition with high accuracy, whose measurement results for the output light power are monitored in a real-time manner and feedback from the detection is delivered to the encoder to update the pumping pattern for self-error correction. Another work \cite{filipovich2022silicon} proposed Silicon photonic architecture that employs the direct feedback alignment training algorithm, which trains neural networks using error feedback rather than error backpropagation, and can operate at speeds of trillions of multiply–accumulate (MAC) operations per second while consuming less than one picojoule per MAC operation. Their system includes MRRs that modulate the incoming laser light with the error vector e and transimpedance amplifiers (TIAs) with tunable gain to convert photocurrent to voltage and scale it to implement the Hadamard product. The authors believe that in the quest for better performance, scalability, and reliability, manufacturing processes and fabrication techniques play a pivotal role and should be the focus of neuromorphic and semiconductor researchers now and in the immediate future.

\subsection{PIC's bottlenecks, remaining challenges, and future directions of neuromorphic photonics}

Left panel of Table 3 first compares PICs to electronic ICs and lists the areas where electronic still dominates photonic systems. It then lists some remaining challenges to be solved and future directions to explore and improve for the PIC in the right panel. From this comparison, we can conclude that despite excelling at areas such as energy consumption, parallelism, and latency, PICs still largely lag behind electronic ICs in many other aspects, most evidently cost, scalability, integratability and footprint. The fact that there are still bulky bench-top implementations in Table 1 is testament that PICs have poor integratability and large footprint. As a matter of fact, most recent state-of-the-art PICs are at the state of ICs almost 60 years ago, when Intel produced the very first batch of computer chips. As a result, we are still decades away from seeing photonic chips actually deployed in our smart phones, laptops, and iPads, and even farther away from having full-blown photonic computers capable of running large AI models such as chatGPT or AlphaGo or solving complex scientific problems such as molecular simulation or finite-element analysis. In other words, the total computing power of PICs (capable of handling few-layer DNNs and less than 10,000s of parameters) at the moment is simply not a rival for electronic ICs (capable of handling DNNs with 100s of layers and trillions of parameters). On the flip side, however, we already have some number of PICs deployed in large-scale data centers and telecommunication systems where integratability and footprint are not of concern. Another major concern of PICs is its cost. As seen in Table 1, the cost to fabricate a photonic chip can be quite high, almost three times more expensive than standard ICs. As a matter of act, cost has always been a critical challenge of silicon photonics that inhibits its widespread adoption in the IT and computing industry. So the industry should focus on reducing the cost by either inventing new materials, revolutionizing the fabrication/integration technologies, or simply expanding scalability. Improved scale and scalability will effectively drive the lowering of cost just as the IC industry has experienced over the past 50 years. Furthermore, latest monolithic integration and SOI techniques could largely improve the integration density and bring down cost. All in all, as the Moore's law comes to an end and the photonic version of the Moore's law begins to take off, we expect to see a considerable improvement in PIC's cost, scalability, integratability, and total computing capacity. PICs will eventually co-exist, if not replace, ICs as the backbone of future computing systems.  

\begin{table}[htbp]
\caption{Left panel: photonic integrated circuits compared to electronic integrated circuits, in terms of cost, scalability, integratability, footprint, computing capacity (throughput) etc.. Right panel: remaining challenges and future directions for photonic devices' research.}
\bigskip
\centering
\rowcolors{1}{yellow}{yellow}
\begin{tabular}{|p{0.25\linewidth} | p{0.25\linewidth} || p{0.33\linewidth} |}
\toprule
\textbf{Electronic IC}  & \textbf{PIC} & \textbf{Challenges and Future Directions of PIC}\\ \hline\hline 
Cost \textit{low} (mature fabrication processes and large-scale production) & Cost \textit{medium} (mature fabrication processes but lack of scale)
& Power Efficiency: Enhancing power efficiency remains a significant challenge. Reducing power consumption is crucial, particularly in mobile devices and data centers.\\ \hline

Scalability \textit{excellent}  & Scalability \textit{good} but yet to be demonstrated
& Quantum Limitations: While quantum devices show immense promise, they also present challenges such as decoherence and error correction.  \\ \hline

Integratability \textit{excellent} (billions of transistors per chip, but coming to an end) & Integratability \textit{poor} (1000s of devices per chip, and continuing to grow) & Sustainability: The electronics industry faces questions of sustainability and responsible sourcing of materials, which are central to achieving carbon neutral and tackling the climate change. The photonics industry should avoid the same pattern by reducing carbon footprint. \\ \hline 

Footprint \textit{small} (nanometer-scale) & Footprint \textit{large} (micro-millimeter scale) & Security: As devices become more connected, ensuring the security and privacy of data remains a critical concern \cite{chen2022effective}.\\ \hline

Total computing capacity \textit{high} (DNN w/ hundreds of layers \& up to trillions of parameters ) & Total computing capacity \textit{low} (few-layer DNN \& up to 10,000s of parameters) & Other directions: photonics-electronics co-packaging, reducing cost of electron-beam lithography, addressing time-consuming FEA or FDTD simulations...\\ \hline

\end{tabular}
\end{table}

\section{Conclusion}
Advances in photonics have catalyzed a transformation in computational technologies, with the integration of optoelectronics onto photonic platforms leading the charge. This integration has facilitated the emergence of PICs, which act as the building blocks for ultra-fast artificial neural networks and are pivotal in the creation of next-generation computational devices. These devices are engineered to address the intensive computational demands of machine learning and AI applications across sectors including healthcare diagnostics, complex language processing, telecommunications, high-performance computing, and immersive virtual environments.

Despite the advancements, conventional electronic systems exhibit limitations in speed, signal interference, and energy efficiency. Neuromorphic photonics, characterized by its ultra-low latency, emerges as a groundbreaking solution, carving out a new trajectory for the advancement of AI and ONNs. This review casts a spotlight on the latest developments in neuromorphic photonic systems from the perspective of photonic engineering and material science, critically analyzing the emergent and anticipated challenges, and mapping out the scientific and technological innovations necessary to surmount these obstacles.

The focus is on an array of neuromorphic photonic AI accelerators, examining the spectrum from classical optics to sophisticated PIC designs. It scrutinizes their operational efficiency, particularly in terms of operations per watt, through a detailed comparative analysis emphasizing key technical parameters. The discussion pivots to specialized technologies such as VCSEL/PCSEL and frequency microcomb-based accelerators, accentuating the latest innovations in photonic modulation and wavelength division multiplexing for effective neural network training and inference.

Acknowledging the current technological barriers in achieving computational efficiencies at the PetaOPs/Watt threshold, the review explores prospective strategies to enhance these critical performance metrics. These include the emerging topological insulators and PCSELs as well as strategies to advance fabrication, system scalability and reliability. The exploration aims to not only chart the current landscape but also to forecast the trajectory of neuromorphic photonics in pushing the frontiers of AI capabilities in the near future. All in all, as the Moore's law comes to an end and the photonic version of the Moore's law begins to take off, we expect to see a considerable improvement in PIC's cost, scalability, integratability, and total computing capacity. PICs will eventually replace ICs as the backbone of future computing systems.

\section*{Acknowledgement}
This work is supported by National Natural Science Foundation of China (NSFC) under Grant No.62174144, Shenzhen Science and Technology Program under Grant No.JCYJ20210324115605016, No.JCYJ20210324120204011, No.JSGG20210802153540017, No.KJZD20230923115114027, and No.JCYJ20220818102214030, Guangdong Key Laboratory of Optoelectronic Materials and Chips under Grant No.2022KSYS014, Shenzhen Key Laboratory Project under Grant No.ZDSYS201603311644527; Longgang Key Laboratory Project under Grant No.ZSYS2017003 and No.LGKCZSYS2018000015; Shenzhen Research Institute of Big Data; Innovation Program for Quantum Science and Technology under Grant No.2021ZD0300701. The authors'd like to thanks Mr. Ceyao Zhang and Dr. Feng Yin for their fruitful discussion on ML and LLMs, and thank Mr. Qi Xin, Pengyu Zhou and Mr. Ping Sun for their fantastic art drawings and CAD modellings, and thank Dr. Hongjie Wang and Mr. Xiaolei Shen for their mentorship and revision tips.\par

\section*{Conflict of interests}
The authors declare no conflict of interests.

\section*{Author contributions}
Z.Z conceived and proposed the writing project. R.L. and Z.Z. designed the structure and outline of the paper. R.L. led and arranged the writing of the whole paper. R.L., Y.G., H.H., Y.Z., S.M. wrote the paper together. S.M. and Y.Z. applied for copyright permissions. Z.Z. supervised and mentored the project. Z.Z. funded the project.

\renewcommand*{\bibfont}{\normalfont\small}

\printbibliography

\newpage
\begin{figure}
  \includegraphics{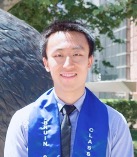}
  \caption*{Renjie Li received his B.S. in Aerospace and Mechanical Engineering from University of California, Los Angeles, CA, USA. In August 2020, he joined the Shenzhen Key Laboratory of Semiconductor Lasers and The Chinese University of Hong Kong, Shenzhen to pursue a doctoral degree. His main research interest is the design and optimization of semiconductor laser devices, machine learning, derivative-free optimization, and LLM.}
\end{figure}

\begin{figure}
  \includegraphics{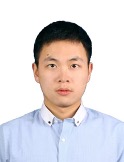}
  \caption*{Yuanhao Gong received his B.Eng. from Northwestern Polytechnical University in Material Science and Engineering, Xi’an, China. In July 2021, he joined the Nano Opto-Electronics Laboratory in the Chinese University of Hong Kong, Shenzhen as a PhD student. His main research interests include the fabrication and growth of micro-cavity lasers on silicon, wafer epitaxy, silicon-on-insulator growth, and photonic crystal surface emitting lasers. }
\end{figure}

\begin{figure}
  \includegraphics{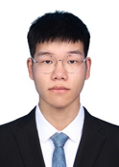}
  \caption*{Hai Huang received his B.Eng. from Harbin Institute of Technology, Harbin, China. Now he is pursuing his PhD at The Chinese University of Hong Kong, Shenzhen. His current research interests mainly involve the design, simulation, and fabrication of PCSEL, TCSEL and other micro-nano laser devices.}
\end{figure}

\begin{figure}
  \includegraphics{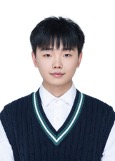}
  \caption*{Yuze Zhou received his B.Eng. in Department of Materials Science and Engineering from Chongqing University, Chongqing, China. In July 2023, he joined the Shenzhen Key Laboratory of Semiconductor Lasers and The Chinese University of Hong Kong, Shenzhen. His main research interest is the design, characterization, and fabrication of semiconductor laser devices.}
\end{figure}

\begin{figure}
  \includegraphics{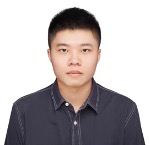}
  \caption*{Sixuan Mao is an undergraduate student in the School of Science and Engineering of the Chinese University of Hongkong (Shenzhen), Shenzhen, China. In June 2023, he joined the Shenzhen Key Laboratory of Semiconductor Lasers. His research interests include nanophotonic devices, LLM, and deep learning.}
\end{figure}

\begin{figure}
  \includegraphics{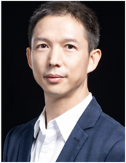}
  \caption*{Prof. Zhaoyu Zhang received his B.S. and M.S. degrees in Applied Mechanics from University of Science and Technology of China, Hefei, China, in 1998 and 2001 respectively. He received Ph.D. degree from California Institute of Technology, Pasadena USA in 2007 in Electrical Engineering. From 2008 to 2011, he worked at University of California, Berkeley as a postdoctoral fellow in the College of Chemistry, with a joint appointment with Lawrence Berkeley National Laboratory. From 2011 to 2015, he worked in Peking University as an Associate Professor and led the team of “Nano-OptoElectronics Lab (NOEL)”. In 2015, he and his team moved to The Chinese University of Hong Kong, Shenzhen. In 2016, he was approved to set up the Shenzhen Key Laboratory of Semiconductor lasers and be the director. His main achievements include the first demonstration of red-emission photonic crystal lasers, wavelength-scale micro-lasers with physical size smaller than 1 micron, microfluidic microlasers based on dye materials, as well as the first demonstration of photonic crystal lasers directly grown on silicon substrates. He has published more than 30 peer-reviewed papers on renowned journals including Nature Communications, Light: Science and Applications, ACS Photonics, Advanced Materials, Nanophotonics, Physics Review Letters, Optica, Photonics research, Optics Letters, Applied Physics Letters, etc. }
\end{figure}

\end{document}